\documentclass[11pt]{article}

\usepackage[utf8]{inputenc}
\usepackage[T1,tone,safe]{tipa}

\usepackage{latexsym,amsmath,amssymb,stmaryrd,theorem,epsfig}

\usepackage{pifont}
\usepackage{array}

\usepackage{setspace}
\usepackage{url}
\usepackage{multirow}
\usepackage[colorlinks=true,linkcolor=blue,urlcolor=blue,citecolor=blue]{hyperref}

\usepackage{mciteplus}
\mciteSetMidEndSepPunct{\nolinebreak;\space}{\nolinebreak.}{\relax}

\DeclareFontFamily{OT1}{rsfs10}{}
\DeclareFontShape{OT1}{rsfs10}{m}{n}{ <-> rsfs10 }{}
\DeclareMathAlphabet{\mathscript}{OT1}{rsfs10}{m}{n}

\numberwithin{equation}{section}

\newcommand{\tr}{\text{tr}}
\newcommand{\cc}[1]{{#1}^*}

\newcommand{\dif}{\mathrm{d}}

\newcommand{\R}{\mathbb{R}}
\newcommand{\diag}{\operatorname{diag}}
\newcommand{\lagr}{\mathcal{L}}
\newcommand{\energy}{\varepsilon}
\newcommand{\JJ}{\textrm{J}}
\newcommand{\blambda}{(1-q^2)}
\newcommand{\hv}{\mathrm{h}}
\newcommand{\al}{\rho}
\newcommand{\alg}[1]{\mathfrak{#1}}

\newcommand{\rf}[1]{(\ref{#1})}

\newcommand{\kp}{b}

\newcommand{\vv}{{\rm v }}
\newcommand{\py}{p}
\newcommand{\Nb}{\text{\!\tiny$N$\!}}

\def \amb {\textrm{c}}
\def \be {\begin{equation}}
\def \ee {\end{equation}}

\def \foot {\footnote}

\def \tr {{\rm tr}}

\def \td {\tilde}
\def \ci {\cite}

\def \ci {\cite}
\def \la {\label}
\def \ep {\energy}
\def \pp {{\rm p}}
\def \hq {{\hat q}}
\def \hh {{\rm h}}
\def \foot {\footnote}
\def \MM {{M}}
\def \iffa {\iffalse}

\def \pe {p}
\def \s {\sigma}
\def \ccc {{\rm c}}

\def \beq {\begin{eqnarray}}
\def \eeq {\end{eqnarray}}
\def \del {\partial}
\def \bea {\begin{align}}
\def \eea {\end{align}}
\def \muu {b}
\def \k {\kappa}
\def \no {\nonumber}
\def \vp {\varphi}
\def \Tp {{\rm T}}

\usepackage[normalem]{ulem}

\begin{document}

\overfullrule=0pt
\parskip=3pt
\parindent=12pt

\vspace{ -3cm}
\thispagestyle{empty}
\vspace{-1cm}

\rightline{ Imperial-TP-AS-2013-01}
\rightline{ HU-EP-13/56}

\begin{center}
\vspace{1cm}
{\Large\bf Giant magnon solution and dispersion relation
\\ \vspace{0.1cm}
in string theory in $AdS_3 \times S^3 \times T^4$ with mixed flux
\vspace{0.1cm} }
\vspace{1.0cm}

{B. Hoare$^{a,}$\footnote{ben.hoare@physik.hu-berlin.de}, A. Stepanchuk$^{b,}$\footnote{a.stepanchuk11@imperial.ac.uk} \ and A.A. Tseytlin$^{b,}$\footnote{Also at Lebedev Institute, Moscow. tseytlin@imperial.ac.uk }}\\

\vskip 0.3cm

{\em $^{a}$ Institut f\"ur Physik, Humboldt-Universit\"at zu Berlin, \\ Newtonstra\ss e 15, D-12489 Berlin, Germany}

{\em $^{b}$ The Blackett Laboratory, Imperial College, London SW7 2AZ, U.K.}

\vspace{.2cm}

\end{center}

\baselineskip 11pt
\begin{abstract}
\noindent
We address the question of the exact form of the dispersion relation for
light-cone string excitations in string theory in $AdS_3 \times S^3\times T^4$
with mixed R-R and NS-NS 3-form fluxes. The analogy with string theory in
$AdS_5 \times S^5$ suggests that in addition to the data provided by the
perturbative near-BMN expansion and symmetry algebra considerations there is
another source of information for the dispersion relation -- the semiclassical
giant magnon solution. In earlier work in arXiv:1303.1037 and arXiv:1304.4099
we found that the symmetry algebra constraints, which are consistent with a
perturbative expansion, do not completely determine the form of the dispersion
relation. The aim of the present paper is to fix the dispersion relation by
constructing a generalisation of the known dyonic giant magnon soliton on $S^3$
to the presence of a non-zero NS-NS flux described by a WZ term in the string
action (with coefficient $q$). We find that the angular momentum of this
soliton gets shifted by a term linear in world-sheet momentum $\pp$. We also
discuss the symmetry algebra of the string light-cone S-matrix and show that
the exact dispersion relation, which should have the correct perturbative BMN
and semiclassical giant magnon limits, should also contain such a linear
momentum term. The simplicity of the resulting bound-state picture provides a
strong argument in favour of this dispersion relation.
\end{abstract}

\newpage

\setcounter{equation}{0}
\setcounter{footnote}{0}
\setcounter{section}{0}

\tableofcontents

\baselineskip 14pt

\section{Introduction}\label{sec:int}

Recently two of us studied superstring theory on $AdS_3\times S^3\times T^4$
with mixed R-R and NS-NS 3-form fluxes \cite{Hoare:2013pma,Hoare:2013ida} with
the aim of solving it using the same integrability-based methods as
developed in the pure R-R flux case (see \ci{Babichenko:2009dk,Borsato:2013hoa}
and references therein). The tree-level light-cone gauge S-matrix for BMN
string excitations \ci{Berenstein:2002jq} was computed in \cite{Hoare:2013pma}.
These excitations have the following perturbative dispersion relation
\begin{align}
\ep_{_\pm} = \sqrt{ 1 - q^2 + (\pe \pm q)^2 } = \sqrt{(1 \pm q \pe)^2+(1-q^2)\pe^2}
\ . \la{1}
\end{align}
Here $ 0 \leq q \leq 1$ is the coefficient of the NS-NS flux ($\hq= \sqrt{ 1 -
q^2}$ is the coefficient of the R-R flux) and $\pe$ is the spatial momentum of
a 2-d string fluctuation.\foot{The quantized coefficient of the WZ term in the
string action is $k= 2 \pi q \hh$.} In general, \rf{1} is expected to receive
corrections at higher orders in the inverse string tension ($\hh = \frac{ \sqrt
\lambda }{ 2 \pi}$) expansion. To obtain the exact S-matrix the first step is
to find the exact generalisation of the dispersion relation \rf{1}. Using
symmetry considerations, as in the $AdS_5 \times S^5$ case \ci{Beisert:2010jr},
in \cite{Hoare:2013ida} the exact generalisation of the dispersion relation was
suggested to be
\begin{align}
\ep_{_\pm} =\sqrt{M^2_\pm +4(1-q^2)\hv^2\sin^2\frac{\pp}{2}}\ , \label{2}
\end{align}
where the ``central charge'' $M_\pm$ is not uniquely determined.
The condition that \rf{2} should reduce to \rf{1} in the near-BMN limit
$ \hh \gg 1, \ \pp \ll 1$ with $\pe = \hh \pp$ fixed implies that
\be
M_\pm = 1 \pm q\hh \pp + \ldots = 1 \pm q\, \pe + O(\hh^{-1}) \ . \la{3}
\ee
If one assumes that the dispersion relation should be manifestly periodic in
$\pp$ (i.e. with $M_\pm$ being a smooth periodic function of $\pp$, which would
apply if there were an underlying spin chain system) then the simplest
consistent form of $M$ would be \ci{Hoare:2013ida}
\be
M_\pm = 1 \pm 2 q \hh \sin \frac{\pp}{2} \ . \la{4}
\ee
As was noted in \ci{Hoare:2013ida}, such a manifestly periodic dispersion
relation \rf{2},\rf{4} suggestive of an underlying spin chain picture also
naturally emerges upon formally discretizing the spatial direction in the
string action (with step $\hh^{-1}$).

There is, however, no a priori reason to expect a spin chain interpretation to
apply to the string integrable system for $q\not=0$. It does not apparently
apply for $q=1$ when the world-sheet theory is related to a WZW model (which is
solved in conformal gauge using, e.g., an effective free-field representation).
For this reason it would be important to have an independent argument for or
against the explicitly periodic choice \rf{4} made in \ci{Hoare:2013ida}.

\

In the $AdS_5 \times S^5$ string case, in addition to the light-cone symmetry
algebra considerations and the perturbative near-BMN expansion, there was a
third string-theory-based source of information for the dispersion relation --
the semiclassical giant magnon solution. The aim of the present paper is to use
this third approach to complement previous work on the first two approaches
\cite{Hoare:2013pma,Hoare:2013ida} and shed further light on the exact form of
the mixed-flux dispersion relation. Following
\ci{Hofman:2006xt,Chen:2006gea,Dorey:2006dq}, we consider a giant magnon
solution on $S^3$ with two angular momenta $(J_1,J_2)$ and find that its energy
is given by ($E,J_1 \to \infty$)
\begin{eqnarray}
&&E-J_1 = \sqrt{ {\MM}^2_\pm + 4(1-q^2)\hh^2\sin^2\frac{\pp}{2}}\ , \label{5}\\
&&
\ \ \ \ \ \MM_\pm = J_2 \pm q\hh \pp \ . \la{6}
\end{eqnarray}
For $q=0$ this reduces to the standard dispersion relation for a dyonic giant
magnon \ci{Chen:2006gea,Dorey:2006dq}. In the giant magnon construction the
momentum $\pp$ is related to the angle $\Delta \phi_1$ between the end-points
of an open rigid string moving along a circle of $S^3$ so that $\pp\in
(-\pi,\pi)$. One may formally consider the energy as periodic in $\pp$ by
periodically extending \rf{6} to the whole interval $\pp \in (-\infty,
\infty)$.\foot{The periodicity in $\pp$ becomes irrelevant in the perturbative
string theory limit of $\hh \gg 1$ when we set $\pp= \hh^{-1} p$ for fixed $p$
so that $\pp$ goes to zero.}

The giant magnon solution is interpreted as a bound state of $J_2$ elementary
``magnons'' (string excitations) so that for $J_2=1$ this relation corresponds
to \rf{2} with an exact linear expression for $M_\pm$ (i.e. without any higher
order corrections in \rf{3})
\be
M_\pm = 1 \pm q\hh \pp\ . \la{7}
\ee
The resulting dispersion relation \rf{5},\rf{7} has the nice feature that for
$q=1$, i.e. in WZW model limit, it directly reduces to the expected massless
dispersion relation
\be
\la{78}
q=1:\qquad \qquad \ep_{_\pm}
= 1 \pm \hh \pp \ .
\ee

To derive \rf{5},\rf{6} we shall start with the bosonic string moving in
$\R\times S^3$ in the presence of an NS-NS flux, i.e. described by an action
with a WZ term proportional to $q$, and consider its classical solutions (see
also \ci{Hoare:2013pma} and references therein). Some previous discussions of
similar classical solutions in this model appeared in
\cite{Lee:2008sk,Chen:2008vc,Huang:2006vz} but they will not be used here.
Since the string model on $\R\times S^3$ in the conformal gauge can be
interpreted as a principal chiral model with a WZ term proportional to $q$, to
find solutions for $q\ne 0$ from solutions in the $q=0$ case one may use the
fact that the $q\ne 0$ equations of motion written in terms of $SU(2)$ currents
are related to the $q=0$ equations of motion through a worldsheet coordinate
transformation.

\

In section \ref{sec:frmodel} we will review the classical string equations on
$\mathbb{R} \times S^3$ in conformal gauge described by the $SU(2)$ principal
chiral model with a WZ term proportional to $q$. We will then discuss the
corresponding conserved charges, pointing out an ambiguity in the action
related to boundary terms, and describe a procedure for constructing classical
solutions for $q\not=0$ from their $q=0$ counterparts, illustrating it on the
example of the rigid circular string solution on $S^3$.

In section \ref{sec:giantMagnon} we will construct the dyonic giant magnon
solution generalising the solution of
\cite{Hofman:2006xt,Chen:2006gea} to the $q\not=0$ case. We will find the
corresponding relation between the energy, the finite angular momentum
component $J_2$, and the effective kink charge, equal to the separation angle
$\Delta \phi_1$ between the rigid open string endpoints. Claiming that the
latter should be interpreted as in \cite{Hofman:2006xt,Chen:2006gea} as the
magnon world-sheet momentum $\pp$, we obtain the dispersion relation
\rf{5},\rf{6}.

In section \ref{sec:llLimit} we will further justify this momentum
identification by considering the limit of large angular momentum which
isolates and effectively decouples fast string motion of extended slowly
varying string configurations such as the giant magnon. In this limit the
string motion is described by a $q\not=0$ generalisation of the familiar
Landau-Lifshitz model
\cite{Kruczenski:2003gt,Kruczenski:2004kw,Kruczenski:2004cn}. The
Landau-Lifshitz equations are known to admit a ``spin wave'' soliton
\cite{TjonWright,Takhtajan:1977rv,Fogedby,Minahan:2008re} which may be
interpreted as the large $J_2$ limit of the dyonic giant magnon solution. The
world-sheet momentum $\pp$ of this Landau-Lifshitz soliton has a
straightforward definition that confirms its identification with $ \Delta
\phi_1$ of the giant magnon. The resulting dispersion relation represents the
large $J_2$ limit of \rf{5}, i.e. 
\begin{align} 
E_{_{\rm LL}} =E-J_1-J_2= -q\hv \pp + \frac{2(1-q^2)\hv^2}{J_2}\sin^2\frac{\pp}{2} + O(J_2^{-2}) \ .
\la{8} \end{align}

In section \ref{sec:5} we will revisit the discussion of the world-sheet
S-matrix of the mixed-flux $AdS_3 \times S^3$ theory in
\cite{Hoare:2013pma,Hoare:2013ida}. We will first review the light-cone
symmetry algebra and then suggest a modification to the conjecture for the
central charge function $M_\pm$ in \cite{Hoare:2013ida}, switching from \rf{7}
to \rf{4}. Doing so, we recover the semiclassical $q\not=0$ dyonic giant magnon
dispersion relation \rf{5},\rf{6} by considering the bound states of elementary
excitations (with $J_2$ being the number of constituents) and taking an
appropriate strong-coupling limit. The simplicity of the bound-state picture
provides a strong argument in favour of the linear momentum function \rf{7}.

Some concluding remarks will be made in section \ref{sec:6}. In appendix
\ref{app:a} we will comment on the relation between the dyonic giant magnon
solution and the soliton of the corresponding Pohlmeyer reduced theory.

\section{\label{sec:frmodel}Classical string solutions on \texorpdfstring{$\R\times S^3$}{R x S3} for \texorpdfstring{$q\ne 0$}{q!=0} }

In this section we shall discuss the relation between the $q=0$ and $q\ne 0$
classical string equations on $\R\times S^3$ that we will use in the following
section to find the unique generalisation of the standard $q=0$ dyonic giant
magnon solution of \ci{Chen:2006gea} to $q\ne 0$. We will see that the $q=0$
and $q\ne 0$ equations written in terms of the current $\alg{J}=g^{-1} dg$ are
related by a worldsheet coordinate transformation. Our strategy will be (i) to
perform this worldsheet coordinate transformation on the $q=0$ current of a
given solution to obtain its $q\ne 0$ counterpart and (ii) starting with this
new current to solve for the coordinates of the $q\ne 0$ solution.

The string action in the conformal gauge is equivalent to that of the principal
chiral model with a Wess-Zumino term with the coefficient $q\in(0,1)$
\begin{align}\label{21}
S = -\frac{\hv}{2}\Big[\int\dif^2\sigma\,\tfrac{1}{2}\tr(\alg{J}_+\alg{J}_-)
- q\int\dif^3\sigma\,\tfrac{1}{3}\varepsilon^{abc}\tr(\alg{J}_a\alg{J}_b\alg{J}_c)\Big]\ ,
\qquad \alg{J}_a = g^{-1}\partial_a g\ ,
\end{align}
where $\hh$ is the string tension, $g\in SU(2)$ and $\sigma^\pm = \frac12 (\tau
\pm \sigma)$,\ $\partial_\pm = \partial_\tau \pm \partial_\sigma$.

\subsection{\label{sec:kk}Classical equations}

The equation of motion for the above action is
\begin{equation}
(1+q)\partial_- \mathfrak{J}_+ + (1-q)\partial_+ \mathfrak{J}_- = 0\ , \qquad \mathfrak{J} = g^{-1} d g\ , \la{23}
\end{equation}
or, equivalently
\begin{equation}
(1-q)\partial_- \mathfrak{K}_+ + (1+q) \partial_+ \mathfrak{K}_- = 0\ , \qquad \mathfrak{K} = dg g^{-1} \ . \la{24}
\end{equation}
Supplemented with the flatness condition \eqref{23} can be rewritten as
\begin{align}
\qquad \partial_+\alg{J}_-+\frac{1}{2}(1+q)\big[\alg{J}_+,\alg{J}_-\big] = 0\ ,
\qquad \partial_-\alg{J}_+-\frac{1}{2}(1-q)\big[\alg{J}_+,\alg{J}_-\big] = 0\ .\label{jEoms}
\end{align}
The formal transformation of the worldsheet coordinates
\begin{align}
\sigma^\pm\rightarrow \tilde \sigma^\pm = (1\pm q)\sigma^\pm \label{wsChange}
\end{align}
then maps the $q\ne 0$ current equations to the $q=0$ equations, provided the
currents are left unaltered (i.e. this is not a conformal transformation that
leaves the classical equations invariant). Furthermore, the Virasoro
conditions (assuming that the target space time coordinate is $t=\kappa \tau$)
\be
\tr(\alg{J}_\pm^2) = -2\kappa^2\ \la{29} \ee
are invariant under this transformation. Given a solution for $q=0$ the map
\eqref{wsChange} allows us to construct the $q\ne 0$ counterpart, $\alg{J}$, of
the $q=0$ current. It then remains to solve the defining equations of
$\alg{J}$ for the function $g$, or, e.g., for the 4 real (2 complex) $S^3$
embedding coordinates $X_m, $ $m=1,...,4$ ($Z_i$, $i=1,2$)
\begin{align}
& \alg{J}_\pm = g^{-1}\partial_\pm g\ ,
\qquad g = \left(\begin{array}{cc}
Z_1 & Z_2 \\ -\cc{Z}_2 & \cc{Z}_1
\end{array}\right) \in SU(2)\ ,\label{jparam}\\
Z_1 & = X_1 + i X_2\ ,\quad Z_2 = X_3 + iX_4\ ,
\qquad X_m^2 = 1\ ,
\qquad \lvert Z_i \rvert^2 = 1\ .
\end{align}
The relation $\partial_\pm g = g\alg{J}_\pm$ then gives first order
differential equations for the complex embedding coordinates in terms of the
current $\alg{J}$
\begin{align}
\partial_\pm{\vec Z} = \alg{J}_\pm^T{\vec Z}\ .\label{JZeq}
\end{align}
Taking an additional derivative these imply
\begin{align}
\partial_+\partial_-{\vec Z} &= (\partial_-\alg{J}_+)^T{\vec Z}+\alg{J}_+^T\alg{J}_-^T{\vec Z}\ ,\qquad
\partial_-\partial_+{\vec Z} = (\partial_+\alg{J}_-)^T{\vec Z}+\alg{J}_-^T\alg{J}_+^T{\vec Z}\ .
\end{align}
Subtracting these equations gives the compatibility condition
$\partial_+\partial_-{\vec Z} = \partial_-\partial_+{\vec Z}$, which corresponds
to the flatness condition for $\alg{J}$. Adding the two equations one
obtains
\begin{align}
\partial_+\partial_-{\vec Z} + \Omega{\vec Z}+\frac{1}{2}q[\alg{J}_+,\alg{J}_-]^T{\vec Z}=0\ ,\qquad \Omega = -\frac{1}{2}\tr(\alg{J}_+\alg{J}_-)\ ,\label{embedEq}
\end{align}
where have we used the fact that since $\alg{J}_\pm$ are traceless and anti-hermitian the
anti-commutator $\{\alg{J}_+,\alg{J}_-\}$ is proportional to the identity.

In order to solve for $\vec Z$ we decouple \eqref{JZeq} into two
first order equations
\begin{align}
\alg{J}_-^{21}\partial_+Z_1-\alg{J}_+^{21}\partial_-Z_1+(\alg{J}_+^{21}\alg{J}_-^{11}-\alg{J}_-^{21}\alg{J}_+^{11})Z_1 &= 0\ ,\label{1stOrdEq1}\\
\alg{J}_-^{12}\partial_+Z_2-\alg{J}_+^{12}\partial_-Z_2+(\alg{J}_+^{12}\alg{J}_-^{22}-\alg{J}_-^{12}\alg{J}_+^{22})Z_2 &= 0\ .\label{1stOrdEq2}
\end{align}
These linear first order partial differential equations can be solved using the
method of characteristics and their solution will involve an undetermined
function. At the same time, the original equations \eqref{JZeq} are four first
order equations for two variables, which uniquely determine the solution up to
integration constants. Therefore, we still need to impose some additional
conditions. This we can do by decoupling the second order equations
\eqref{embedEq} as
\begin{align}
\partial_+\partial_-Z_1 +q\frac{C_{12}}{\alg{J}_+^{21}}\partial_+Z_1
+\Big(\Omega +q C_{11}-q\frac{C_{12}}{\alg{J}_+^{21}}\alg{J}_+^{11}\Big)Z_1 = 0\ ,\label{embedDec1}\\
\partial_+\partial_-Z_2 +q\frac{C_{21}}{\alg{J}_+^{12}}\partial_+Z_2
+\Big(\Omega +q C_{22}-q\frac{C_{21}}{\alg{J}_+^{12}}\alg{J}_+^{22}\Big)Z_2 = 0\ ,\label{embedDec2}
\end{align}
where $ C= \frac{1}{2}[\alg{J}_+,\alg{J}_-]^T$. The undetermined function can
then be fixed by substituting the solution of the first order equations into
the above second order equations.\footnote{Notice that the equations
\eqref{embedDec1} and \eqref{embedDec2} are related through complex
conjugation. This does not imply that $Z_1$ and $Z_2$ are complex conjugates of
each other as we shall see in more detail later on. The above second order
differential equations admit two separate solutions corresponding to roots of a
quadratic equation. Requiring that the final solution is consistent with the
original $q=0$ solution then uniquely fixes the choice of these roots giving
rise to solutions for $Z_1$ and $Z_2$ that in general are not related by
complex conjugation.}

\subsection{\label{sec:cc}Conserved charges}

The equations of motion \rf{23},\rf{24} imply that we have two conserved
$SU(2)$ currents, the left-invariant and the right-invariant one
\begin{equation}
L_a = \mathfrak{J}_a - q \epsilon_{ab} \mathfrak{J}^b\ , \qquad R_a = \mathfrak{K}_a + q \epsilon_{ab}\mathfrak{K}^b\ , \qquad \partial_a L^a = \partial_a R^a = 0\ .
\end{equation}
Using these we can construct conserved charges in the standard way
\begin{equation}\label{charges}
Q_L = \hv\int \textrm{d} \sigma \, (\mathfrak{J}_0 + q \mathfrak{J}_1) \ , \qquad
Q_R = \hv\int \textrm{d} \sigma \, (\mathfrak{K}_0 - q \mathfrak{K}_1) \ .
\end{equation}
For the dyonic giant magnon solution that is the main subject of this paper,
there are a particular pair of charges that we will be interested in
\begin{equation}
\JJ = -\frac i4(\textrm{Tr}[Q_L\cdot \sigma_3] + \textrm{Tr}[Q_R\cdot \sigma_3])\ , \qquad
\MM = -\frac i4(-\textrm{Tr}[Q_L\cdot \sigma_3] + \textrm{Tr}[Q_R\cdot \sigma_3])\ . \la{25}
\end{equation}

We will parametrise the 3-sphere as
\begin{align}
Z_1 = X_1 + i X_2 = \sin\theta\ e^{i\phi_1}\ ,\qquad Z_2 = X_3 + i X_4 = \cos\theta\ e^{i\phi_2}\ ,\label{hopf}
\end{align}
or, equivalently, in terms of the Euler angles
\begin{align}
g = e^{\frac{i}{2}\theta_L\sigma_3}e^{\frac{i}{2}\psi \sigma_2}e^{\frac{i}{2}\theta_R\sigma_3}\ ,
\qquad \psi =\pi- 2\theta\ ,\quad \theta_L = \phi_1+\phi_2\ ,\quad \theta_R = \phi_1-\phi_2\ . \la{eu}
\end{align}
The bosonic string action \rf{21} then takes the form
\be
{S} = -\frac{\hh}{2} \int\dif^2\sigma\, \Big[\partial_a \theta \partial^a \theta + \sin^2\theta \partial_a \phi_1 \partial^a \phi_1 + \cos^2\theta \partial_a \phi_2 \partial^a \phi_2 + q(\cos 2\theta + \amb) (\dot{\phi}_1\acute{\phi}_2-\dot{\phi}_2\acute{\phi}_1)\Big]\ ,\label{123} \quad
\ee
where $a,b = 0,1$ stand for the worldsheet coordinates $\tau$, $\sigma$ with
the metric $\eta=\diag(-1,1)$ and $\dot{}=\partial_\tau$,
$\acute{}=\partial_\sigma$. The last $q$-dependent term comes from the
Wess-Zumino term, in which the parameter $\amb$ corresponds to an ambiguity in
defining a local 2-d action.\foot{In general, the string couples locally to 
the antisymmetric $B$-field, while the defining equations -- the conformal
invariance conditions or, to leading order, the supergravity equations of
motion -- depend on the three-form field strength $H$. Therefore, there is a
gauge freedom in choice of the $B$-field and in the presence of the boundary
this necessitates a boundary term parametrising this ambiguity. At the moment
we will leave it arbitrary and fix it later via natural physical requirements
appropriate for the giant magnon solution.} The $\amb$-term is a total
derivative and does not, of course, affect the equations of motion. However, if
we consider string solutions with non-trivial boundary conditions (which
includes the case of interest -- the dyonic giant magnon) then it will affect
the corresponding Noether global charges as we shall discuss below.

As we are using the conformal gauge with the residual conformal symmetry fixed
by choosing $t=\kappa \tau$ the Virasoro constraints take the following
explicit form
\begin{equation}
\dot{\theta}^2 +\acute{\theta}^2 + \sin^2\theta(\dot{\phi}_1^2+\acute{\phi}_1^2)+\cos^2\theta(\dot{\phi}_2^2+\acute{\phi}_2^2) = \kappa^2\ ,
\qquad \dot{\theta}\acute{\theta}+\sin^2\theta \dot{\phi}_1\acute{\phi}_1+\cos^2\theta \dot{\phi}_2\acute{\phi}_2 = 0.
\end{equation}
The translational invariance of the full string action under shifts of $t$,
$\phi_1$ and $\phi_2$ leads to the following conserved Noether charges: the
energy and the angular momenta (here $\sigma \in (-\pi , \pi)$)
\begin{align}
E &= 2\pi \hv\kappa\ ,\qquad \la{222} \\ J_1 &= \hv
\int\dif \sigma\,\big[\sin^2\theta \dot{\phi}_1-\frac{q}{2}(\cos2\theta+\amb)\acute{\phi}_2\,\big]\ ,\la{223} \\
J_2 &= \hv
\int\dif \sigma\,\big[\cos^2\theta\dot{\phi}_2+\frac{q}{2}(\cos2\theta+\amb)\acute{\phi}_1\,\big]\ .\label{224}
\end{align}
Here $J_{1,2}$ follow directly from the action \eqref{123}. The $\amb$-terms
are of course total derivatives and thus contribute only if $\phi_1$ or
$\phi_2$ have non-trivial boundary values or if periodicity in $\s$ is not
imposed (as will be the case for the giant magnon solution).

Let us compare \rf{223},\rf{224} with the charges $\JJ$ and $\MM$ \eqref{25}
that were derived from the $SU(2)$-invariant currents \rf{charges}.
Substituting the parametrisation \eqref{hopf} into \eqref{charges} we find
\begin{align}
\JJ &= \hv
\int\dif \sigma\,\big[\sin^2\theta \dot{\phi}_1-\frac{q}{2}(\cos2\theta+1)\acute{\phi}_2\,\big]\ ,\la{225} \\
\MM &= \hv
\int\dif \sigma\,\big[\cos^2\theta\dot{\phi}_2+\frac{q}{2}(\cos2\theta-1)\acute{\phi}_1\,\big]\ ,\label{226}
\end{align}
and hence
\beq
&& J_1 = \JJ - \frac12 \hh q ( \ccc-1) \Delta \phi_2 \ , \ \ \ \ \ \ \ \
J_2 = \MM - \frac12 \hh q ( \ccc+1) \Delta \phi_1\ , \la{227} \\
&& \ \ \ \ \ \qquad \qquad \Delta \phi_i = \phi_i(\pi) - \phi_i(-\pi)\ . \la{228}
\eeq
Thus to match $\JJ$ and $M$ with $J_1$ and $J_2$ we need different choices of
$\amb$ $(=\pm 1)$, i.e. $\JJ$ and $M$ cannot be obtained as Noether charges
from a local action \eqref{123} with equations of motion equivalent to \rf{23}.
This, of course, is not a contradiction as the difference appears only due to
the boundary ``twist'' terms $\Delta \phi_i $, but if non-zero such terms break
manifest $SU(2)$ symmetry.\foot{Let us note also that in general, the currents
conserved ($\del_a j^a_i =0$) on the equations of motion are defined modulo a
trivial term $\epsilon^{ab} \del_b f_i$ where the functions $f_i$ (that may, in
principle, break some manifest global symmetries) contribute to the
corresponding charges only if they have non-trivial boundary twists.}

The dyonic giant magnon solution we will be interested in is a classical
soliton representing a ``bound state'' of string excitations above the BMN
vacuum. The latter corresponds to a point-like string moving along a great
circle of $S^3$
\begin{equation}
\theta = \frac\pi 2 \ , \ \ \ \ \ \ \phi_1 = \kappa \tau\ ,\ \ \ \ \phi_2 =0 \ . \la{ 230}
\end{equation}
For the point-like BMN solution
\begin{equation} \
E-J_1 = 0\ , \ \ \ \ \ \ \ \ \ J_1 =\JJ \ . \la{231}
\end{equation}
In the $q=0$ case, the giant magnon limit \cite{Hofman:2006xt} involves taking
both $E$ and $J_1$ to infinity (i.e. $\kappa \to \infty$) with their difference
held fixed
\begin{equation}\label{hmlimit}
E,\ J_1 \to \infty, \qquad \qquad
\epsilon\equiv E- J_1, \ J_2 \ = \ \textrm{fixed}\ .
\end{equation}
Also, as in \cite{Hofman:2006xt} the string is assumed to be open so that
rescaling $\tau$ and $\sigma$ by $\kappa\to \infty $ the spatial interval may
be decompactified
\be
x=\kappa \s \ , \ \ \ \ \ \ \ \kappa \to \infty \ ,\ \ \ \ \ \ x \in ( -\infty, +\infty)\ , \la{ka}
\ee
and the non-zero angle between the end points of the string
\begin{equation}\label{topcharge}
\Delta \phi_1 = \phi_1 (x = \infty) - \phi_1(x = -\infty)\ ,
\end{equation}
may be related to the 2-d momentum $\pp$. Then $\epsilon$, which plays the
role of the energy of the state relative to the BMN vacuum, can be expressed as
a function of $\pp$ and $J_2$.

As we shall see below, for $q\not=0$ the requirement that $E-J_1$ remains
finite in the $\kappa \to \infty$ limit (and also the classical action is
finite when evaluated on one period of the dyonic giant magnon solution)
implies that
\be \la{233}
\amb =1\ . \ee
In this case the action \rf{123} becomes explicitly
\be
{S} = -\frac{\hh}{2} \int\dif^2\sigma\, \Big(\partial_a \theta \partial^a \theta + \sin^2\theta\, \partial_a \phi_1 \partial^a \phi_1 + \cos^2\theta\, \partial_a \phi_2 \partial^a \phi_2 + 2q\epsilon^{ab} \cos^2 \theta\, \del_a \phi_1 \del_b \phi_2 \Big)
\ . \label{1230} \quad
\ee
The physical reason for this particular choice of $B_{mn}$-term in the string
action is that it vanishes at $\theta = \frac{\pi}{2}$. This implies the
vanishing of force on the ends of the open string (representing the giant
magnon solution) moving along the great circle corresponding to $\phi_1$. As
usual, the boundary term in the variation of the string action specifies the
boundary conditions for the open string end-points. The variation of \rf{1230}
under the variation of $\phi_1$ gives the condition
$\int d\tau\ \delta \phi_1 \big(\sin^2 \theta \del_\sigma \phi_1 - q \cos^2 \theta \del_\tau \phi_2 \big)\Big|_{\sigma=0, \pi} =0$.
Since the end-points of the giant magnon move along the great circle
$\theta\big|_{\sigma=0, \pi}=\frac{\pi}{2}$ the $q$-dependent term vanishes and
we just have the standard free-ends condition $\del_\sigma \phi_1
\big|_{\sigma=0, \pi}=0$.

From \rf{228} we then get 
\begin{equation} \la{234} \JJ = J_1 \ , \ \ \ \ \ \ \ \ \
\MM
= J_2 - q\hv \Delta \phi_1\ .
\end{equation}
Here $\Delta \phi_1$ plays the role of kink charge, which, as for $q=0$, can be
identified with the 2-d spatial momentum $\pp$ of the soliton. Recalling that
the quantized WZ level $k$ is related to $q$ as $k = 2\pi \hh q$, we may write
$\MM$ as
\be \la{236}
\MM = J_2 - q\hv \pp = J_2 - k \frac{\pp}{ 2 \pi} \ , \ \ \ \ \ \ \qquad
\pp=\Delta \phi_1 \ . \ee
Here $ \Delta \phi_1 \in (0, \pi) $ but being an angular coordinate it may
defined modulo $2 \pi$, and then the same may be assumed about $\pp$, i.e.
$\MM$ may be considered as a periodic function of $\pp$.\footnote{The issue of
periodicity is a subtle one and we will return to it later in section
\ref{sec:dr} after we have derived the relevant expressions for the energy and
angular momenta of the dyonic giant magnon as functions of the solution
parameters (which include $\Delta\phi_1$).}

Also, for a physical closed string $\Delta \phi_1$ should be equal to $2\pi n$
where $n$ is an integer winding number, or, equivalently, the total momentum of
a bound state of magnons representing a physical state should be quantized
\be \sum_i \pp_i = 2 \pi n \ . \ee
This is consistent with both $\MM$ and $J_2$ in \rf{236} taking integer values
for such states.

This relation between $\MM$ and $J_2$ is suggestive of how the dyonic giant
magnon dispersion relation is to be modified in the presence of the NS-NS flux
(cf. \rf{5},\rf{6}).

\subsection{\label{sec:rgCircStrings}An example of a solution: rigid circular string}

Before turning to the construction of the giant magnon solution for $q\not=0$
let us illustrate the general procedure of finding $q\not=0$ solutions from
their $q=0$ counterparts on the example of a rigid circular string on $S^3$
\ci{Frolov:2003qc,Arutyunov:2003za}. The standard $q=0$ solution written in
the embedding coordinates reads
\begin{align}
Z_1 &= \frac{1}{\sqrt{2}}\exp[i(\omega+m)\sigma^++i(\omega-m)\sigma^-]\ ,
\nonumber\\
Z_2 &= \frac{1}{\sqrt{2}}\exp[i(\omega-m)\sigma^++i(\omega+m)\sigma^-]\ ,\qquad
\qquad m^2+\omega^2 = \kappa^2\ . \la{237}
\end{align}
For this solution the $SU(2)$ currents are (see \rf{jparam})
\begin{align}
\alg{J}_+ &= i\left(\begin{array}{cc}
m & \omega \exp[{-2im(\sigma^+-\sigma^-)}] \\
\omega \exp[{2im(\sigma^+-\sigma^-)}] & -m
\end{array}\right)\ ,\no \\
\alg{J}_- &= i\left(\begin{array}{cc}
-m & \omega \exp[{-2im(\sigma^+-\sigma^-)}]\\
\omega \exp[{2im(\sigma^+-\sigma^-)}] & m
\end{array}\right)\ ,\no \\
\qquad \Omega &= -\frac12 \tr (\alg{J}_+ \alg{J}_-) = \omega^2-m^2\ .
\end{align}
Performing the worldsheet coordinate transformation \eqref{wsChange} gives the
$q\ne 0$ currents
\begin{align}
\alg{J}_+ = i\left(\begin{array}{cc}
m & \omega \exp[{-2im[\sigma^+-\sigma^-+q(\sigma^++\sigma^-)]}]\\
\omega \exp[{2im[\sigma^+-\sigma^-+q(\sigma^++\sigma^-)]}] & -m
\end{array}\right)\ ,\no \\
\alg{J}_- = i\left(\begin{array}{cc}
-m & \omega \exp[{-2im[\sigma^+-\sigma^-+q(\sigma^++\sigma^-)]}]\\
\omega \exp[{2im[\sigma^+-\sigma^-+q(\sigma^++\sigma^-)]}] & m
\end{array}\right)\ ,\no \\
C = 2m\omega\left(\begin{array}{cc}
0 & \exp[2im(\sigma^+-\sigma^-+q(\sigma^++\sigma^-))]\\
-\exp[-2im(\sigma^+-\sigma^-+q(\sigma^++\sigma^-))] & 0
\end{array}\right)\ ,
\end{align}
with $\Omega$ unchanged, while the decoupled equations for the embedding
coordinates \eqref{embedDec1}-\eqref{embedDec2} become
\begin{align}
\partial_+\partial_-Z_1 -2qmi\,\partial_+Z_1 +\Big(\omega^2-m^2 -2qm^2\Big)Z_1 &= 0\ ,\\
\partial_+\partial_-Z_2 +2qmi\,\partial_+Z_2 +\Big(\omega^2-m^2 -2qm^2\Big)Z_2 &= 0\ .
\end{align}
Fourier decomposing the solution as
\begin{align}
Z_1 = a_n\exp[i\mu_n\tau+in\sigma]\ ,\qquad Z_2 = b_n\exp[i\nu_n\tau-in\sigma]\ ,
\end{align}
and requiring that the modes reduce to those of the $q=0$ circular string
solution one obtains
\begin{align}
\mu_m = qm + \sqrt{q^2m^2+\omega^2}\ , \qquad \nu_m = -qm + \sqrt{q^2m^2+\omega^2}\ .
\end{align}
The normalisation condition $\lvert Z_1\rvert^2+\lvert Z_2\rvert^2=1$ together
with the Virasoro constraints $\partial_\pm Z_1\partial_\pm\cc{Z}_1 +
\partial_\pm Z_2\partial_\pm \cc{Z}_2=\kappa^2=\omega^2+m^2$ then determine
$a_n$ and $b_n$ up to a phase giving
\begin{align}
Z_1 & = \sqrt{\frac{W-qm}{2W}}\exp(i[(W+q m)\tau+m\sigma])\ ,\la{248}
\\ Z_2 & = \sqrt{\frac{W+qm}{2W}}\exp(i[(W-q m)\tau-m\sigma])\ ,\la{249}
\qquad W = \sqrt{\omega^2+q^2m^2}\ .
\end{align}
In the parametrisation \eqref{hopf} this solution takes the form
\begin{align}
\sin\theta = \sqrt{\frac{W-qm}{2W}}\ ,
\qquad \phi_1 = & (W+qm)\tau+m\sigma\ , \qquad \phi_2 = (W-qm)\tau-m\sigma\ .\label{circSolQne0}
\end{align}
The two angular momenta associated to shifts in $\phi_1$ and $\phi_2$ computed
from \rf{223},\rf{224} are
\begin{align}
J_1 &= J_2 = \pi \hh (W+\amb qm) \ ,
\end{align}
where $\amb$ parametrises the ambiguity in the choice of the total derivative
term in the action \rf{123}. The expression for energy then takes the form
\begin{align}\la{252}
E &= 2\pi \hh \kappa = \sqrt{(J- 2\pi \hh \amb q m)^2 + 4 \pi^2 \hh^2 m^2(1-q^2)}\ ,\qquad \qquad J\equiv J_1+J_2 = 2 J_1 \ .
\end{align}
Expanding in large $J$ we get
\begin{align}
E &= J- 2 \pi \hh \amb qm
+\frac{2 \pi^2 \hh^2(1-q^2) m^2}{J}+O(J^{-2})\ .\label{rgCircStringE}
\end{align}
The choice $\ccc=0$ here gives the standard BMN limit $E=J$ when $J\rightarrow
\infty$.

\section[Dyonic giant magnon on \texorpdfstring{$\mathbb{R} \times S^3$}{R x S3} in the presence of NS-NS flux]{\label{sec:giantMagnon}Dyonic giant magnon on \texorpdfstring{$\R\times S^3$}{R x S3} in the presence of NS-NS flux}

\subsection{Review of \texorpdfstring{$q=0$}{q=0} case}

Let us start with a review of the standard dyonic giant magnon solution on
$S^3$ in the absence of an NS-NS flux, i.e. for $q=0$ in the action \rf{123}.
In the notation of section \ref{sec:frmodel} the dyonic giant magnon solution,
labelled by the two independent parameters $\kp$ (or $v$) and $\al$, takes the
form \cite{Chen:2006gea}
\begin{align} \label{30}
&Z_1 = \frac{ \big[\kp + i \tanh({\cal X}\cos\al) \big]\ \exp(it)}{(1+\kp^2)^{1/2}}\ , \qquad \qquad
Z_2 = \frac{\mathrm{sech}({\cal X}\cos\al)\ \exp(i{\cal T}\sin\al)}{(1+\kp^2)^{1/2}}\ ,\\
&
\kp = \frac{v\, \sec\al}{ \sqrt{1-v^2}}\ , \qquad \qquad v\in(0,1)\ , \qquad \rho \in (0, \frac{\pi}{2})\ , \qquad b \in (0,\infty) \ ,
\la{31}
\end{align}
where ${\cal X}$ and ${\cal T}$ are related to the worldsheet coordinates
$\tau$, $\sigma$ through a boost of velocity $v$ and a rescaling by $\kappa$
\begin{align}\la{32}
{\cal X} = & \frac{x-vt}{\sqrt{1-v^2}}\ , \qquad\qquad {\cal T} = \frac{t-vx}{\sqrt{1-v^2}}\ ,
\\ t = & \kappa\tau,\ \ \ x= \kappa\sigma \ ,\qquad \tau\in (-\infty,\infty)\ ,
\qquad \sigma\in (-\pi,\pi)\ , \ \ \ \ \ \ \ \ \ x \in (-\infty,\infty) \ . \la{33}
\end{align}
Here we have already taken the limit $\k\to \infty$ and thus ``decompactified''
the spatial direction $x$. $x\rightarrow\pm\infty$ correspond to the endpoints
of the string moving in the $\phi_1$ direction, while
$\al\in(0,\frac{\pi}{2})$ is the parameter associated with the angular momentum
in the $\phi_2$ direction. We may of course extend the parameter ranges so
that $v \in (-1,1), \ \rho \in
(-\pi,\pi)\setminus\{-\frac{\pi}{2},\frac{\pi}{2}\}$ to cover also the soliton
moving in the opposite direction. These ranges correspond to $b \in
(-\infty,\infty)$ and hence $\Delta\phi_1 \in (-\pi,\pi)$.

This solution satisfies the boundary conditions
\begin{align}
x\rightarrow \pm \infty : \ \ \ \ \ \ \ \ Z_1\rightarrow \exp\Big(it\pm i\frac{\Delta\phi_1}{2}\Big)\ ,
\qquad Z_2\rightarrow 0\ , \label{GMbcs}
\end{align}
where
\begin{align}
{ \Delta\phi_1} = 2\arctan \kp^{-1}\ \ \in (0, \pi) \label{366}
\end{align}
corresponds to the angle between the rigid open string endpoints which move in
the $\phi_1$ direction on the great circle $\theta = \frac\pi 2$.

The finite combination of energy $E$ with $J_1$ and the angular momentum $J_2$
for this solution are given by
\begin{align}
E-J_1 = \frac{2\hv}{1+\kp^2}\frac{(1+\kp^2\cos^2\al)^{1/2}}{\cos\al}\ ,\qquad\qquad
J_2 = \frac{2\hv}{1+\kp^2}\tan\al\ .\la{36}
\end{align}
The case of $\rho=0$ thus corresponds to the $S^2$ giant magnon of
\ci{Hofman:2006xt} (then $Z_2$ in \rf{30} becomes real) with $J_2=0$.
In addition to $J_2$ another ``charge'' parameter of this solution is the
``kink charge'' $ \Delta\phi_1$. Expressing the energy in terms of these
charges we get
\begin{align}
E-J_1 = \sqrt{J_2^2+\frac{4\hv^2}{1+\kp^2}}= \sqrt{J_2^2+4\hv^2\sin^2\frac{\Delta\phi_1}{2}}\label{dDisp} \ .
\end{align}
This becomes the usual dyonic giant magnon dispersion relation upon the
identification \ci{Hofman:2006xt,Chen:2006gea} of the magnon momentum with the
separation angle: $\pp=\Delta\phi_1$.

Let us mention that if one considers a more general solution where the string
moves also along an $S^1$ in the torus part of $AdS_3 \times S^3 \times T^4$
then the dispersion relation \rf{dDisp} is modified as follows: 
\begin{align}
\sqrt{ E^2 - P^2}-J_1 = \sqrt{J_2^2+4\hv^2\sin^2\frac{\pp}{2}}\label{3666} \ ,
\end{align}
where $P$ is the (large) momentum in $S^1$ with $E,P$ and $J_1$ scaling as
$\kappa \to \infty$. This follows simply from the formal Lorentz invariance in
the $\mathbb{R}_t \times S^1{}_{\!\!\!\psi}$ subspace in the decompactification
limit (equivalently, the contribution of the $\mathbb{R}_t$ and
$S^1{}_{\!\!\!\psi}$ to the Virasoro condition can be absorbed into a rescaling
of $\tau$ and $\sigma$).

\subsection{\label{sec:32}Dyonic giant magnon for \texorpdfstring{$q\not=0$}{q!=0}}

Let us now generalise the above solution to the $q\not=0$ case using the procedure
outlined in section \ref{sec:frmodel}. First we are to re-express the current,
constructed from the $q=0$ solution \eqref{30} via \eqref{jparam}, in terms of
$\tilde\sigma^\pm$, defined in \eqref{wsChange}, giving us the current
$\alg{J}$ of the $q\ne 0$ solution. Anticipating that the $q\ne 0$ solution is
again most conveniently written in terms of boosted worldsheet coordinates we
introduce the boosted worldsheet coordinates $\tilde {\cal X}$, $\tilde {\cal
T}$ which are related to the $q\ne 0$ worldsheet coordinates $\tilde t$,
$\tilde x$ by a boost of velocity $v$, i.e.
\be
\tilde\sigma^\pm =(1 \pm q) \sigma^\pm = \frac{1}{2}(\tilde\tau\pm \tilde\sigma)\ ,
\qquad \tilde t = \kappa\tilde\tau\ ,\quad \tilde x = \kappa\tilde\sigma\ ,
\qquad \tilde{\cal X} = \frac{\tilde x-v\tilde t}{\sqrt{1-v^2}}\ ,\quad \tilde{\cal T}=\frac{\tilde t-v\tilde x}{\sqrt{1-v^2}}\ .\la{38}
\ee
Note that the boosted worldsheet coordinates of the $q=0$ and $q\ne 0$ cases
are related via
\begin{align}
\tilde{\cal X} = {\cal X} + q{\cal T}\ ,\qquad\qquad \tilde{\cal T} = {\cal T}+q{\cal X}\ .\la{xxx}
\end{align}
It is useful also to introduce rescaled coordinates $\tilde\xi$ and
$\tilde\eta$, defined as
\be
\tilde\xi = \tilde{\cal X}\cos\al\ ,\qquad\qquad \tilde\eta = \tilde{\cal T}\sin\al\ . \la{39}
\ee
The coordinate transformation from the original light-cone coordinates
$\sigma^\pm$ to $\tilde \xi$, $\tilde\eta$ is then
\begin{equation}
\sigma^+ = \frac{1}{2 \k (1+q)}\sqrt{\frac{1+v}{1-v}}\Big(\frac{\tilde\eta}{\sin\al}+\frac{\tilde\xi}{\cos\al}\Big)\ ,
\qquad \sigma^- = \frac{1}{2\k(1-q)}\sqrt{\frac{1-v}{1+v}}\Big(\frac{\tilde\eta}{\sin\al}-\frac{\tilde\xi}{\cos\al}\Big)\ .\la{310}
\end{equation}
Written in $\tilde \xi$, $\tilde\eta$ coordinates the equations
\eqref{1stOrdEq1}-\eqref{1stOrdEq2} become
\begin{equation}
\partial_{\tilde\xi} Z_1 + A\,\partial_{\tilde\eta} Z_1 + B\, Z_1 = 0\ ,
\qquad \partial_{\tilde\xi} Z_2 + \cc{A}\,\partial_{\tilde\eta} Z_2 + \cc{B}\, Z_2 = 0\ ,\label{XTdiffEqs}
\end{equation}
where
\begin{align}
A & = \tan\al\,\frac{(1+q)\sqrt{\frac{1-v}{1+v}}\alg{J}_-^{21}-(1-q)\sqrt{\frac{1+v}{1-v}}\alg{J}_+^{21}}{(1+q)\sqrt{\frac{1-v}{1+v}}\alg{J}_-^{21}+(1-q)\sqrt{\frac{1+v}{1-v}}\alg{J}_+^{21}}
= \tan\al\,\frac{m+q\kp+iq \tanh\td \xi}{qm+\kp+i \tanh\td \xi}\ , \\
B & = \frac{\kappa^{-1}\cos^{-1}\al(\alg{J}_+^{21}\alg{J}_-^{11}-\alg{J}_-^{21}\alg{J}_+^{11})}{(1+q)\sqrt{\frac{1-v}{1+v}}\alg{J}_-^{21}+(1-q)\sqrt{\frac{1+v}{1-v}}\alg{J}_+^{21}}
= -i\,\frac{\mathrm{sech}^2\xi+(k+i\tanh\td \xi) (\kp+i u \tanh\td \xi)}{(u+1) (\kp+qm+i\tanh\td \xi )}\ , \\
& \qquad u = \frac{\sqrt{1+\kp^2\cos^2\al}}{\sin\al}\ ,\qquad\qquad m=(1-u)\tan\al\ .
\end{align}
We can then write \eqref{XTdiffEqs} as an ordinary differential equation
\begin{align}
\frac{\dif Z_1}{\dif{\tilde\xi}} + B Z_1 = 0\label{ode1}
\end{align}
valid along the characteristic curve
\begin{align}
\frac{\dif\tilde\eta}{\dif\tilde\xi} = A(\tilde\xi)\ \ \ \ \Rightarrow\ \ \ \ \tilde\eta = \int\dif\tilde\xi\,A(\tilde\xi)+C_0\ .\label{curveEq}
\end{align}
Evaluating the integrals of $A$ and $B$ we obtain
\begin{align}
I_1 = & \int\dif\tilde\xi\, B(\tilde\xi) = -\frac{2i[k+k^2s_1+(k-s_1)u]\tilde\xi}{2(1+s_1^2)(1+u)}+\ln\cosh\tilde\xi\notag\\
& \hspace{63pt} -\frac{[1+k^2+s_1(s_1-k)(1+u)]\ln[2(s_1\cosh\tilde\xi+i\sinh\tilde\xi)^2]}{2(1+s_1^2)(1+u)}\ ,\notag\\
I_2 = & \int\dif\tilde\xi\,A(\tilde\xi) = \tan\al\,\frac{2(1+s_1s_2)\tilde\xi-im(1-q^2)\ln[2(s_1\cosh\tilde\xi+i\sinh\tilde\xi)^2]}{2(1+s_1^2)}\ ,\notag\\
& \qquad s_1 = k + qm\ ,\qquad \qquad s_2 = m+ qk\ .
\end{align}
The solution for $Z_1$ is then obtained by integrating \eqref{ode1}
\begin{align}
Z_1 = f(C_0(\tilde\xi,\tilde\eta)) \exp\Big[-\int\dif\tilde\xi\, B(\tilde\xi)\Big]
=f(\tilde\eta-I_2(\tilde\xi))\exp\big[-I_1(\tilde\xi)\big]\ .\label{30General}
\end{align}
We can determine $f$ by substituting this solution into \eqref{embedDec1}. This gives
\begin{align}
f''(x) - 2r f'(x) + r^2-\delta^2 &= 0\ ,\\
r \equiv \frac{i}{2}\Big(\frac{1}{\sin\al\sqrt{1-v^2}}-1\Big)\ ,\qquad & \qquad
\delta \equiv\frac{i}{2}\Big(1+\frac{1+q(q-2v)}{\sin\al(1-q^2)\sqrt{1-v^2}}\Big)\ ,
\end{align}
which has the solutions
\begin{align}
f(x) = e^{a_\pm x}\ ,\qquad a_\pm = r\pm\delta\ .
\end{align}
Requiring that in the limit $q\rightarrow 0$ we recover the dyonic giant magnon
solution \eqref{30} leads to
\begin{align}
f(z)=e^{a_+z}= \exp\Big(i\frac{z}{\sin\al\sqrt{1-v^2}}\frac{1-qv}{1-q^2}\Big)\ .
\end{align}
We can now determine the $Z_2$ solution by taking the complex conjugate of
\eqref{30General}, but to ensure the correct $q=0$ limit in this case we should
take
\begin{align}
f(z)=e^{a_-z}= \exp\Big(iz\Big[\frac{q(q-v)}{\sin\al(1-q^2)\sqrt{1-v^2}}+1\Big]\Big)\ .
\end{align}
After fixing the normalisation constants using the Virasoro condition and
$\lvert Z_1\rvert^2+\lvert Z_1\rvert^2=1$ we obtain
the solution written in terms of the original ${\cal X}$, ${\cal T}$ coordinates \rf{32}
\begin{align}
&Z_1 = \frac{ \big( \tilde \kp + i \tanh[ \cos\al({\cal X}+q{\cal T})] \big)\ \exp (i t) }{(1+\tilde \kp^2)^{1/2}}\ , \la{323} \\
&Z_2 = \frac{\mathrm {sech}[\cos\al({\cal X}+q{\cal T})]\ \exp\big(i\big[ \sin\al({\cal T}+q{\cal X})-qx\big]\big)}{(1+\tilde \kp^2)^{1/2}}\ ,\la{324} \\
&\tilde \kp = \sec \al\big(\frac{v-q}{\sqrt{1-v^2}}+q \sin \al\big) \ .
\label{325}
\end{align}
This generalises \rf{30},\rf{31} to the $q\not=0$ case. 
It is straightforward to verify that the solution \rf{323}--\rf{325} satisfies
the defining equations \eqref{JZeq}. Written in the parametrisation
\eqref{hopf} it takes the form\foot{One can check that this solution remains
valid also for $q=1$ (it satisfies the Virasoro constraints and equations of
motion for \eqref{123}) even though the worldsheet coordinate transformation
\eqref{wsChange} which we used to derive it becomes degenerate.
Furthermore, written in terms of the group element \rf{eu} the solution factorises as
expected: \ $g = \exp\big(\frac i2(t-x)\,\sigma_3\big)\cdot g_{_R}(t+x)$. It is
interesting to note that the right-invariant current is particularly simple in
this limit: \ $\partial_- g g^{-1} = i\,\sigma_3$.}
\begin{align}
\cos\theta = \frac{\mathrm{sech}[\cos\al\ ({\cal X}+q{\cal T})]}{(1+\tilde \kp^2)^{1/2}}\ , & \label{326} \\
\phi_1 = t+\arctan\big(\tilde \kp^{-1}\tanh[\cos\al\ ({\cal X}+q{\cal T})]\big)\ , & \qquad
\phi_2 = \sin\al\, ({\cal T}+q{\cal X})-qx\ , \label{327}
\end{align}
where as in \rf{32} here ${\cal X} = \frac{x-vt}{\sqrt{1-v^2}}\ ,\ \ {\cal T} =
\frac{t-vx}{\sqrt{1-v^2}}$.

The asymptotics of this $q\ne 0$ dyonic giant magnon solution \rf{323},\rf{324}
have the same form as in the $q=0$ case \rf{GMbcs}
\begin{align}
x\to \pm \infty :\ \ \ \ \ \ & Z_1\rightarrow \exp\Big(it\pm i\frac{\Delta\phi_1}{2}\Big)\ ,
\qquad Z_2\rightarrow 0\ , \label{GMbcs1}\\
&
\Delta\phi_1 = 2\arctan \tilde\kp^{-1}\label{329} \ \in(0,\pi) \ .
\end{align}
Here we have restricted so that $\Delta \phi_1 \in (0,\pi)$, corresponding to
$\tilde \kp \in (0,\infty)$. As in the $q=0$ case these ranges can be extended
to $(-\pi,\pi)$ and $(-\infty,\infty)$ respectively.

\subsection{\label{sec:dr}Conserved charges and dispersion relation}

For the $q=0$ dyonic giant magnon the energy $E$ and the angular momentum $J_1$
diverge with their difference staying finite. This is no longer true in general
for $q\ne 0$: the behaviour of $E-J_1$ happens to depend on the definition of
$J_1$ in \rf{223} which is sensitive to the total derivative ambiguity $(\sim
\ccc$) in the Wess-Zumino term in \eqref{123}.\footnote{The infinite
contribution of the total derivative term comes from the infinite (in the
$\k\to \infty$ limit) number of ``windings'' of the $q\ne 0$ giant magnon
around the circle of $\phi_2$, i.e. $\Delta \phi_2 = \infty$ in \rf{327}.} We
find from \rf{222}--\rf{224}
\begin{align}
E- J_1 &= \hv\int_{-\infty}^{\infty}\dif x\,\Big( 1-\big[
\sin^2\theta\partial_t\phi_1-\frac{q}{2}(\cos2\theta+\amb)\partial_x\phi_2\big]\Big)\ ,\la{3291}\\
J_2 &= \hv\int_{-\infty}^{\infty}\dif x\,\Big[\cos^2\theta\partial_t\phi_2+\frac{q}{2}(\cos2\theta +\amb)\partial_x\phi_1\Big]\ ,\la{330}
\end{align}
where we used the rescaled worldsheet coordinates
$(t,x)=(\kappa\tau,\kappa\sigma)$ with $t,x\in (-\infty,\infty)$.
Computing these integrals for the solution \rf{326},\rf{327} we find
\begin{align}\label{334}
E- J_1 & = 2\hv\frac{\sqrt{1-q^2+(\tilde \kp\cos\al-q\sin\al)^2}}{(1+\tilde \kp^2)\cos\al}
+\frac12 \hh q (\ccc-1) \Delta \phi_2 \ ,\\
\Delta \phi_2 &= - \cos\al\ (q\cos\al+\tilde\kp\sin\al) \ x\big|_{{-\infty}}^{{\infty}} \ , \la{3344}
\\
J_2 &= \MM + \frac12 (\amb + 1) \hh q \Delta \phi_1 \ ,
\qquad \ \ \ \ \ \ \
\MM = 2 \hh \frac{ \tan\al-q\tilde \kp }{1+\tilde \kp^2}
\ , \label{335}
\end{align}
where $\Delta \phi_1$ is given in \rf{329}, the divergent expression for
$\Delta \phi_2 = \phi_2(x=\infty) - \phi_2(x=-\infty) $ follows from \rf{327} and
$M$ was defined in \rf{25},\rf{226}. We conclude that $E- J_1$ is finite only
if $\ccc=1$.\foot{Let us note again that this choice is not related to the
parameters of the solution itself but only to the total derivative term in the
action \eqref{123} or to the definition of the corresponding Noether charge
$J_1$. Let us mention also that the importance of similar WZ-term related
boundary terms in the presence of non-trivial kinks was emphasised in a similar
context in \ci{Hollowood:2011fm}.} Remarkably, this is exactly the case (cf.
\rf{227}) when the charge $J_1$ \rf{223} coincides with $\JJ$ in
\rf{25},\rf{225} which corresponds to manifestly $SU(2)$ invariant current.

Eliminating $\al$ and expressing $\td \kp$ in terms of $\Delta \phi_1$ in \rf{329} gives
\be \la{3355}
\amb=1 :\ \ \ \ \ \ \ \ \ \ \ E-J_1 = \sqrt{M^2+4\hv^2(1-q^2)\sin^2\frac{\Delta\phi_1}{2}} \ , \ \ \ \ \ \ \ \ \ \ \ \ \ \ \ \ \
M= J_2- q\hh \Delta\phi_1 \ .
\ee
Let us comment on the values of parameters here (with $q\in (0,1)$). As in the
$q=0$ case, when constructing the solution we restrict to $\Delta \phi_1 \in
(0,\pi)$, or equivalently $\td \kp \in (0,\infty)$. Taking also $\rho \in
(0,\frac\pi 2)$, this implies the restriction $v > v_*(q,\rho) > v_*(q,0) = q$,
where $v_*$ is a function of $q$ and $\rho$ whose explicit form follows from
\rf{325}. As before, we may extend the parameter ranges so that $v \in (-1,1)$
and $\al \in (-\pi,\pi)\setminus\{-\frac\pi2,\frac\pi2\}$ and thus $\tilde \kp
\in (-\infty,\infty)$ and $\Delta\phi_1 \in (-\pi,\pi)$.

Note also that $\MM$ in \rf{335} is single-valued; $J_2= \MM + q\hv\Delta\phi_1
$ formally shifts if we shift $\Delta \phi_1$ by its period. As was already
mentioned in section \ref{sec:cc}, the shift is integer as the WZ level $k=2
\pi \hh q$ should be quantized.

It remains to relate the ``kink charge'' $\Delta\phi_1$ \rf{329} to the
world-sheet momentum $\pp$. In general, there is no universal definition of
world-sheet momentum (the total momentum vanishes as we are dealing with a
reparametrisation-invariant theory). In the present case the preferred gauge
used to define the near-BMN S-matrix is the uniform light-cone gauge (see
\ci{Arutyunov:2009ga} for a review). In the $q=0$ case the momentum was
identified in \cite{Hofman:2006xt,Chen:2006gea} with the angular separation
\begin{equation}\label{momid}
\pp = \Delta\phi_1\ ,
\end{equation} and this
relation was indeed demonstrated to apply in the uniform light-cone gauge for
the original ($J_2=0$) \cite{Hofman:2006xt} giant magnon
\ci{Arutyunov:2006gs,Arutyunov:2009ga}. Heuristically, the relation \rf{momid}
is not expected to change upon switching on non-zero values of $J_2$ and $q$.
First, there should be no momentum flow in $\phi_2$ direction in \rf{327} as it
is linear in $t$ and $x$, i.e. the relevant momentum should be associated with
$\phi_1$. Expressing $\phi_1$ and $\phi_2$ in terms of two other string
coordinates -- $t$ and $\theta$ in \rf{326},\rf{327} and treating $\theta$ as a
spatial coordinate along the string ($\cos \theta$ changes from its maximal
value to zero and then back) we get\footnote{This form of the solution is also
useful for understanding its qualitative features. In particular, we see that
the three parameters $\tilde b$, $w$ and $r$ each control a different type
of behaviour. $\tilde b = \cot \frac{\Delta \phi_1}{2} \in (-\infty,\infty)$
measures the distance between the end points of the string, while $w \in
(-1,1)$ measures the angular velocity in the $\phi_2$ direction. The string also
winds in the $\phi_2$ direction and the size of these windings is controlled by
$r \in (-\infty,\infty)$. From the expressions for $\tilde b$, $w$ and
$r$ in \eqref{325},\eqref{343},\eqref{377} it is clear that the NS-NS flux
does not introduce any new qualitative behaviours, i.e. they are all present
for $q=0$. However, for fixed $q$ (i.e. for a given string background) the
solution is parametrised by only two independent parameters (for example
$(v,\rho)$ or $(\tilde b,\rho)$) and hence only certain combinations of these
three behaviours are allowed. As we let $q$ vary the NS-NS flux can support
certain configurations that would not otherwise be obtainable.}
\begin{align} \la{397}
&\phi_1(t,\theta) = t +\arctan\big[\, { \td \kp^{-1} \sqrt{1 - (1+\tilde \kp^{2}) \cos^2\theta}}\, \big]\, \ , \\
& \la{3977}
\phi_2 
(t,\theta) = w t + r \, { \rm arccosh} \big[ \, \big(\sqrt{ 1 + \tilde \kp^2} \cos \theta\big)^{-1} \big] \ , \\
&w =\frac{(1-q^2) \sin \rho\, \sqrt{1- v^2}-q (v-q)}{ 1-qv } = \frac{\sqrt{1-q^2+ (\tilde b \cos\rho - q\sin\rho)^2 }}{\sin\rho - q \tilde b \cos\rho} \ , \qquad \Tp \equiv \frac{2\pi}{|w|} \ , \la{343}
\\ & r = w \frac{q + \tilde b \tan \rho } { q \tilde b - \tan \rho } \ . \la{377} 
\end{align}
Here the independent parameters are $\rho$ and $\tilde b$ associated, respectively, 
with two conserved charges -- $J_2$ and $\pp$ (see \rf{329},\rf{335}). 
The expression for $\phi_1$ has indeed the same form as for the $J_2=0, \ q=0$
case, i.e. it depends on $\rho$ (or $J_2$) and $q$ only via $\td b$ in
\rf{325}. Then a natural definition of the world-sheet momentum corresponding
to $\phi_1$ is
\be \la{398}
\pp = \int d \theta \ \del_t \phi_1 \, \del_\theta \phi_1 = \int d \theta \ \del_\theta \phi_1 = \Delta \phi_1 \ ,
\ee
where we have taken into account that $\del_t \phi_1(t, \theta) =1$.\foot{Let
us note also that general, given a rigid moving-wave soliton described by some
profile function $\vp=\vp(x- v t) $ one may define the momentum as $\pp= \int d
x \, p_\vp\, \vp' $ where $p_\vp$ is the momentum density corresponding to $f$.
For the $S^2$ giant magnon described in the light-cone gauge with constant
$J_1$-density, this leads to $\pp = 2 \arccos v$ \ci{Arutyunov:2006gs}. One
can see that the relation $\cos\frac{\pp}{2} = \vv$ between $\pp$ and the
soliton center of mass velocity $\vv$ generalises also to $J_2\not=0$ and
$q\not=0$ cases ($\vv=v$ in \rf{30} when $\rho=0$). From \rf{323},\rf{324} the
string centre of mass coordinates are
$z_i = \lim_{\kappa\rightarrow\infty}\int_{-\pi}^{\pi} \frac{\dif \sigma}{2\pi} Z_i(t,\k \s)$, 
i.e. $z_1 = \frac{\tilde\kp}{\sqrt{1+\tilde\kp^2}} \, e^{it}$, \ $z_2 = 0$, 
i.e. they describe a motion along a circle in the $(X_1,X_2)$ plane
with linear (tangent) velocity given by (cf. \rf{329}) \ $\vv =
\frac{\tilde\kp}{\sqrt{1+\tilde\kp^2}} = \cos\frac{\Delta\phi_1}{2}$.} The
same conclusion is indeed reached in the uniform light-cone gauge where one has
\ci{Arutyunov:2009ga} \ $x_- = \vp- t$, \ $x_+ = (1-a) t + a \vp = \tau$, \ $p_+
=(1-a) p_\vp -a p_t =1$, \ $p_- = p_\vp + p_t$. Here $a$ is a gauge parameter
(we ignore winding in the $\vp$ direction as we are interested in the
decompactification limit $J_1 \to \infty$). The Virasoro condition (which is
unchanged by the presence of the WZ term $\sim q$) then implies $ \acute{x}_- +
p_i \acute{x}^i =0$. In the present case $\vp$ is to be identified with
$\phi_1$ (see \ci{Hoare:2013pma}) and ${x}^i$ stand for all other
``transverse'' coordinates. Thus the world-sheet momentum is $p_{\rm ws}
\equiv - \int d\sigma\ p_i \acute{x}^i = \int d\sigma\ \acute{x}_- = \Delta \vp
=\Delta\phi_1=\pp$.\foot{Thus the relation between world-sheet momentum and
$\Delta \phi_1$ does not depend on the gauge parameter $a$; this was observed
for $q=0$ in \ci{Arutyunov:2006gs} and agrees also with the near-BMN expansion
for $q\not=0$ in \ci{Hoare:2013pma}.}

Using \rf{momid} in \rf{3355} we arrive at the following $q\ne 0$
generalisation of the dyonic magnon dispersion relation\footnote{For 
completeness,  one should also check that the ``off-diagonal'' components of the
$SU(2)$ charges \eqref{charges} vanish on this solution and indeed this is the
case.
We can also  construct these ``off-diagonal'' charges by considering the
corresponding Noether currents following from the local action \eqref{123}. For
these charges to be well-defined, i.e. for the spatial component of the current
to go to zero as $x \rightarrow \pm \infty$ we again find that we should fix
$\ccc = 1$.  Furthermore, these charges also vanish on the solution, i.e. there
is no additional contribution from the non-trivial boundary conditions.}
\begin{align}
E-J_1 = \sqrt{(J_2-q\hv \pp)^2+4\hv^2(1-q^2)\sin^2\frac{\pp}{2}}\ .\label{340}
\end{align}
It is worth noting that, as in the $q=0$ case (see equation \eqref{3666}), the
generalisation of this dispersion relation to the case when the string also
moves along an $S^1$ in the torus part of the background is simply given by
replacing $E \to \sqrt{E^2-P^2}$ where $P$ is the (large) momentum in $S^1$.
Again the reason for this is the formal Lorentz invariance in the $\mathbb{R}_t
\times S^1{}_{\!\!\!\psi}$ subspace in the decompactification limit.

\

Finally, let us derive the quantization condition for $J_2$ (a similar argument
for the $q=0$ case appeared in \cite{Hofman:2006xt,Chen:2006gea}). As one can
see from \rf{324} or \rf{3977} the giant magnon motion is time-periodic in the
$\phi_2$ direction with period $\Tp$ \rf{343} assuming that the shift of $t$ is
compensated by a shift of $x$ so that ${\cal X}+q{\cal T}$ and thus $\theta$
stays unchanged. In fact, the solution is explicitly periodic in $x_+ =\tau$ in
the light-cone gauge discussed above, where $x_+ =(1-a)t + a \phi_1 = t + a x_-
= \tau $ ($x_-=f(\theta)$, $\phi_2= w \tau + g(\theta)$, see
\rf{397},\rf{3977}). The changes over the period $\delta t =\Tp$ are $\delta
\theta =0, \ \delta \phi_2 = 2 \pi $, $\delta \phi_1 =\delta t =\Tp$ so that
$\delta x_- = \delta \phi_1 - \delta t=0 , \ \ \delta x_+= \Tp. $

This periodicity implies that there is an associated action variable, which
should take integer values upon semiclassical quantization. Indeed, in
general, for an integrable Hamiltonian system one can define action
variables $ I_s = \frac{1}{2\pi}\int_{\gamma_s} p_i \dif q^i $ where the
$\gamma_s$ form a basis of Liouville torus cycles. The Bohr-Sommerfeld
condition then implies that $I_s$ should take integer values in the quantum
theory. In the present case we can obtain the action variable $I$ associated
to the above cycle in phase space from\footnote{Note that $\int p_i \dif q^i
= \int_0^\Tp \dif t\int_{-\infty}^{+\infty}\dif x\, p_i \dot{q}^i = S + \Tp H
= S - \Tp \frac{\del S}{\del \Tp} $ where we used the Hamilton-Jacobi equation $
H=-\frac{\partial S(t)}{\partial t } $.} 
\begin{align}\la{b2} 
2 \pi I = S - \Tp \frac{\del S}{\del \Tp}\Big|_\pp \ . 
\end{align}
Here $S=S(\Tp,\pp)$ is the light-cone gauge string action computed over one
period $\Tp$ on the giant magnon solution (we assume that the parameters $\rho$
and $\tilde b$ are expressed in terms of $\Tp$ in \rf{343} and $\pp$ in
\rf{momid},\rf{329}). Since the string action is reparametrisation-invariant,
its value is gauge-independent and so it can be evaluated, e.g., in the
conformal gauge (even though the periodicity of the solution is not manifest in
this gauge -- $\phi_1$ gets an additional shift $\sim \Tp$). Considering $w>0$
we compute the action \rf{123} (keeping $\ccc$ arbitrary and including the
$-\del^a t \del_a t$ term) on the solution \rf{326},\rf{327} to find
\begin{align}
S = 2\pi \hh \Big[ - \frac{2(1-q^2)}{\tan\rho- q\tilde \kp} + \frac{1}{2}q(\ccc+1)\Delta\phi_1 - \frac{1}{4 \pi} \Tp q(\ccc-1)\Delta\phi_2\Big]\ , 
\end{align}
where $\Tp$ is given in \eqref{343}, $\Delta\phi_1$ in \rf{329},\rf{momid}, and
$\Delta\phi_2$ in \rf{3344} is divergent. Thus the action, like $E-J_1$ in
\rf{334}, is finite only if $\ccc=1$, once again supporting the choice of the
boundary term made above in \rf{233},\rf{1230}. Eliminating $\rho$ in favour
of $\Tp$ or $w= \frac{2 \pi}{\Tp}$ (recall we consider $w>0$) using
\eqref{343}, i.e.
\begin{align}
\tan^2\rho = \frac{1-q^2+\tilde\kp^2-\Big[\sqrt{(1-q^2)(1+\tilde\kp^2)(1-w^2)}-qw\tilde \kp\Big]^2}{1-w^2} \ ,
\end{align}
gives (here $\tilde \kp = \cot \frac{\pp}{2} $ and $w = \frac{2\pi}{\Tp}$)
\begin{align}
\ccc=1:\ \ \ \ \ \ \frac{S}{2\pi \hh} = 
\frac{2(1-q^2)\sqrt{1-w^2}}{q\tilde\kp\sqrt{1-w^2}-\Big(1-q^2+\tilde\kp^2-\Big[\sqrt{(1-q^2)(1+\tilde\kp^2)(1-w^2)}-qw\tilde \kp\Big]^2\Big)^{1/2}} + q \pp \ . 
\end{align}
Substituting into \rf{b2} we find that the action variable associated to the
periodic motion in $\phi_2$ is nothing but $J_2$ given in \rf{335}, i.e.
\be
I = J_2 \ . \la{ij}
\ee
Thus $J_2$ should be quantized, which is consistent with the near-BMN
perturbation theory \ci{Hoare:2013pma} where the dispersion relation is a limit
of \rf{340} with $J_2 = 1$, and with the bound-state analysis in section
\ref{sec:5}.

\section{\label{sec:llLimit}Giant magnon in the Landau-Lifshitz limit}

Let us now check \rf{momid},\rf{340} by
considering a particular large angular momentum limit (when both $J_1$ and $J_2$ are large) in which the string action
reduces to a Landau-Lifshitz (LL) model in which there is a natural definition for
the world-sheet momentum.

 In the $q=0$ case the LL model admits
a well-known ``spin wave'' soliton solution which, in fact, may be interpreted
as a limit of the giant magnon of the original string sigma model, and we shall
find its generalisation to $q\not=0$. In the LL model one can give a natural
definition to the spatial 2-d momentum of the soliton and as we shall see it is
consistent with \rf{momid},\rf{398} and the resulting energy-momentum relation
agrees with large $J_2$ expansion of \rf{340}.

\subsection{\label{41}Landau-Lifshitz model for \texorpdfstring{$q\not=0$}{q!=0}}

To derive the LL model from the string action on $\mathbb{R} \times S^3$ one
introduces a collective coordinate to isolate the ``fast'' string motion
associated to the large total angular momentum and obtains the effective action
describing the remaining ``slow'' degrees of freedom
\cite{Kruczenski:2003gt,Kruczenski:2004kw,Kruczenski:2004cn}. Let us
parametrise $S^3$ as
\begin{align}
Z_1 &= X_1+iX_2 = \sin\theta\, e^{i\phi_1} = U_1e^{i\alpha}\ ,\qquad U_1 = \sin\theta\, e^{i\beta}\ ,\nonumber\\
Z_2 &= X_3+iX_4 = \cos\theta\, e^{i\phi_2} = U_2e^{i\alpha}\ ,\qquad U_2 = \cos\theta\, e^{-i\beta}\ ,\nonumber\\
\alpha &= \frac{1}{2}(\phi_1+\phi_2)\ ,\qquad \beta = \frac{1}{2}(\phi_1-\phi_2)\ ,\qquad \lvert U_1\rvert^2+\lvert U_2\rvert^2=1\ .\label{s1hopfs3}
\end{align}
The angle $\alpha$ and the $\mathbb{C}\mathbf{P}^1$ coordinates $U_1$, $U_2$
correspond to the $S^1$ Hopf fibration of $S^3$. The conformal-gauge string
Lagrangian is $\lagr = -\frac{1}{2}\partial_+t\,\partial_-t +
\frac{1}{2}\lagr_S$ where the $S^3$ part in \rf{123} written in the above
coordinates takes the form
\begin{align}
\lagr_S & = \partial_+\theta\partial_-\theta + \partial_+\alpha\partial_-\alpha +\partial_+\beta\partial_-\beta \notag\\
& \hspace{54pt}-(1+q)\partial_+\alpha C_- -(1-q)\partial_-\alpha C_+ \notag
-q \amb(\partial_+\alpha\partial_-\beta-\partial_+\beta\partial_-\alpha)\ ,\nonumber\\
& C_\pm = \cos2\theta\,\partial_\pm\beta\label{lagrHopf} \ .
\end{align}
With $t=\k \tau$ the Virasoro constraints are
\begin{align}
(\partial_\pm\alpha)^2-2\partial_\pm\alpha C_\pm +(\partial_\pm\theta)^2+(\partial_\pm\beta)^2 = \kappa^2\ .
\end{align}
Introducing $n_i = U^\dagger\sigma_i U$ or explicitly
\begin{align}
&{\vec n} = (\sin 2\theta\,\cos 2\beta,\,\sin 2\theta\,\sin 2\beta,\,\cos 2\theta)\ , \qquad {\vec n}^2 = 1\ ,
\\ &
\partial_+C_--\partial_-C_+ = -\frac{1}{2}\varepsilon_{ijk}n_i\partial_+n_j\partial_-n_k\ ,\qquad
\frac{1}{4}\partial_+{\vec n}\cdot\partial_-{\vec n} = \partial_+\theta\partial_-\theta+\partial_+\beta\partial_-\beta-C_+C_-\ ,
\end{align}
we may rewrite the Lagrangian in \rf{123} and the Virasoro constraints as
\begin{align}
& \lagr_S = \frac{1}{4}\partial_+n\cdot\partial_-n + (\partial_+\alpha - C_+)(\partial_-\alpha-C_-)
-q (\partial_+\alpha C_- -\partial_-\alpha C_+)
\nonumber
\\ & \qquad \qquad \qquad \qquad \, -q \amb (\partial_+\alpha\partial_-\beta-\partial_+\beta\partial_-\alpha)\ ,\label{46}\\
& \partial_\pm\alpha -C_\pm = \kappa\sqrt{1-\frac{(\partial_\pm n)^2}{4\kappa^2}} \ . \label{47}
\end{align}
Let us now take the large total angular momentum limit directly in the action
(as in \cite{Kruczenski:2004kw}) using the Virasoro constraints to eliminate
$\alpha$.\foot{Here we take the limit directly in the action rather than the equations of motion in order to determine the contribution to the LL action from the total derivative in the WZ term.} Introducing $u = \alpha - t$ and expanding in large $\kappa$ (which
corresponds to large angular momentum limit with both $J_1$ and $J_2$ being
large\foot{This one can see from the large $\kappa$ expansion of $J_\alpha = 2(J_1+J_2) = \kappa + \int\frac{\dif\sigma}{2\pi}\,q[\amb +\cos(2\theta)]\acute{\beta} + O(\kappa^{-1})$.}) we find, after solving for $u$ using the Virasoro constraints
$\partial_\pm u = C_\pm + O(\kappa^{-1})$,\footnote{We dropped the total
derivative term $2\kappa\partial_\tau u$ and the constant term $\kappa^2$ which
do not depend on $q$.}
\begin{align}
\lagr_S &=\frac{1}{4}\partial_+ n\cdot\partial_-n - 2\kappa C_\tau +2q \kappa C_\sigma +2q \amb \kappa \partial_\sigma\beta + O(\kappa^{-1})\ .
\end{align}
Finally, using the equation of motion $\partial_\tau n_i=q\partial_\sigma n_i +
O(\kappa^{-1})$ we arrive at the following $q\not=0$ generalisation of the
Landau-Lifshitz action
\begin{align}
S_{_{\rm LL}} &= -\hv\int\dif t\dif x\,\Big[ C_t -q C_x+\frac{1}{8}(1-q^2)(\partial_x n_i)^2 -q \amb \partial_x\beta\Big]\ ,\label{410}
\end{align}
where $t=\kappa\tau$, $x=\kappa\sigma$ and $C_a= \cos 2\theta\ \partial_a
\beta$.\foot{In
this procedure we use the Virasoro constraints in the action which in general
may not necessarily lead to a correct result but in the present case indeed
gives the same expression for the LL action as the systematic procedure
based on uniform gauge fixing and large $\kappa$ expansion developed in
\cite{Kruczenski:2004kw}. The same conclusion is also easily reached for $q\ne
0$ by taking the same limit directly at the level of string equations of
motion:
$$\partial_+n_i(\partial_-\alpha-C_-)+\partial_-n_i(\partial_+\alpha-C_+) -\varepsilon_{ijk}n_j\partial_+\partial_-n_k
+q(\partial_+\alpha\partial_-n_i-\partial_-\alpha\partial_+n_i)+O(\kappa^{-1})=0\ .$$
Using the expansion of the Virasoro constraints \rf{47} to eliminate $\alpha$
gives
$$2\kappa (\partial_\tau-q\partial_\sigma)n_i +
\varepsilon_{ijk}n_j(\partial_\sigma^2-\partial_\tau^2)n_k
+q(C_+\partial_-n_i-C_-\partial_+n_i)+O(\kappa^{-1}) = 0 \ .$$ For $q\not=0$
the time derivatives of ${ n_i}$ are not suppressed but go as $\partial_\tau
n_i = q\partial_\sigma n_i+O(\kappa^{-1})$. Eliminating them recursively from
the above equation gives $(\partial_\tau-q\partial_\sigma)n_i =
-\frac{1}{2\kappa}(1-q^2)\varepsilon_{ijk}n_j\partial_\sigma^2
n_k+O(\kappa^{-2})\ $ which follows from the action \rf{410}.}

The LL model action \rf{410} is invariant under translations of $(t,x)$ and
$SO(3)$ rotations of $n_i$. The former give two conserved charges -- 2-d
energy and 2-d momentum of the ``slow'' variables (which are no longer fixed by
the Virasoro constraints)
\begin{align}
E_{_{\rm LL}} = &
\hv\int\dif x\Big(
-q(\cos 2\theta+ \amb)\partial_x\beta+\frac{1}{2}(1-q^2) \big[(\partial_x \theta)^2+\sin^2 2\theta\, (\partial_x\beta)^2\big]\Big)\ ,\la{411} \\
P_{_{\rm LL}} = &
-\frac{\hv}{2}\int\dif x\,\frac{\partial\lagr_S}{\partial (\partial_t{\beta})}\partial_x\beta = -\hv\int\dif x\,\cos 2\theta\, \partial_x \beta\ .\la{412}
\end{align}
Then
\be
E_{_{\rm LL}}+q P_{_{\rm LL}} = \hv\int\dif x\,\Big( -q\amb\partial_x \beta + \frac{1}{2}(1-q^2)\big[(\partial_x \theta)^2+\sin^2 2\theta\, (\partial_x \beta)^2\big]
\Big)\ . \la{414}
\ee
Before discussing the LL model counterpart of the giant magnon solution let us
consider the corresponding LL limit of the rigid circular string solution of
section \ref{sec:rgCircStrings}. The solution in \rf{248},\rf{249} may be
written as
\begin{equation}
\cos\theta = \frac{1}{\sqrt{2}}\sqrt{1+\frac{qm}{\sqrt{\kappa^2-(1-q^2)m^2}}}\ ,\qquad \alpha = \sqrt{\kappa^2-(1-q^2)m^2}\ \tau\ ,\qquad \beta = m(\sigma+q\tau)\ .
\end{equation}
Taking the large $\k$ limit we get, to leading order,
\begin{align}
\cos\theta = \frac{1}{\sqrt{2}}\ ,\qquad \alpha = \big[\kappa-\frac{(1-q^2)m^2}{2\kappa}\big]\tau\ ,\qquad \beta = m(\sigma+q\tau) \ ,
\end{align}
which solve the LL equations of motion.
The corresponding
conserved charges \rf{411},\rf{412} are
\begin{align}
E_{_{\rm LL}} = 2\pi \hv [- q \amb m+\frac{1}{2}(1-q^2)\frac{m^2}{\kappa}]\ ,\qquad { P}_{_{\rm LL}}=0\ ,
\end{align}
which is consistent with a large angular momentum expansion of the full string
energy \eqref{rgCircStringE} (here $E_{_{\rm LL}} = E- J_1$ and $J_1= 2 \pi \hh
\k$).

\subsection{Landau-Lifshitz limit of the dyonic giant magnon solution}

When taking the Landau-Lifshitz or large $\kappa$ limit we required that the
derivative $\partial_\sigma n_i$ stays finite. However, since the giant magnon
solution is itself a large $\kappa$ solution depending only on
$(t,x)=(\kappa\tau,\kappa\sigma)$, in this case $\partial_\sigma n_i \sim
O(\kappa)$. Therefore we also need to take an appropriate limit of the
parameters in \rf{323}--\rf{325} to obtain the corresponding solution of the LL
model. We have from \rf{326},\rf{327}
\begin{align}
&\partial_\sigma \cos\theta = -\frac{\kappa\cos\al}{\sqrt{1+\tilde\kp^2}}\frac{1-qv}{\sqrt{1-v^2}} \frac{\tanh(\cos\al({\cal X}+q {\cal T}))}{\cosh(\cos\al({\cal X}+q {\cal T}))}\ ,\nonumber\\
&\partial_\sigma \beta = \kappa\cos\al\frac{1-qv}{\tilde\kp\sqrt{1-v^2}}\frac{\cos^2(\arctan(\tilde\kp^{-1}\tanh[\cos\al({\cal X}+q {\cal T})]))}{\cosh(\cos\al({\cal X}+q {\cal T}))}
+\frac{\kappa}{2}\big(q-\sin\al\frac{q-v}{\sqrt{1-v^2}}\big)\ ,
\end{align}
so that to take the LL limit we need to assume that in the large $\kappa$ limit
\begin{align}
\sin\al \sim 1 + O(\kappa^{-2})\ ,\qquad
\qquad v \sim O(\kappa^{-1})\ .
\end{align}
In this limit we have (cf. \rf{31},\rf{325})
\begin{align}
\tan\al \sim \kappa\ \gg 1 \ , \ \ \ \ \ \ \ \ \ \ \ \ \ \muu = \frac{v\sec \rho}{\sqrt{1-v^2}} ={\rm fixed} \ , \ \ \ \ \ \ \ \td b = b + O(\kappa^{-1})\ .
\end{align}
Under this assumption the conserved charges \rf{3291},\rf{330} of the $q\ne 0$
dyonic giant magnon take the form ($\ccc=1$)
\begin{align}
&E-J_1 = \frac{2\hv\kappa}{1+\muu^2} + O(\kappa^0)\ ,\qquad J_2 = \frac{2\hv\kappa}{1+\muu^2} + O(\kappa^0)\ ,\qquad \Delta\phi_1=2\arctan\muu^{-1} + O(\kappa^{-1})\ ,\nonumber \\
&\frac{E-J_1}{J_2} = 1-q\frac{(1+\muu^2)\arctan \muu^ {-1}}{\kappa} + O(\kappa^{-2})\ . \label{gmKappaexpansion2}
\end{align}
Thus in the Landau-Lifshitz limit $E-J_1$ and $J_2$ diverge with their ratio
staying finite. Eliminating $\muu$ and $\kappa$ from the above expressions we
reproduce the large $J_2$ expansion of \eqref{340}
\begin{align}
E-J_1 = J_2-q\hv\Delta\phi_1 + O(J_2^{-1})\ .
\end{align}

To construct the corresponding LL solution let us first consider the $q=0$
case. Expanding the $q=0$ giant magnon \eqref{30} at large $\k$ we
get\foot{Note also that the Virasoro constraints take the expected form
$\partial_\pm\alpha - C_\pm = \kappa - \frac{1}{2\k} \mathrm{sech}^2(\sigma-\frac{\muu}{\kappa}\tau).$}
\begin{align}
2\alpha &= -\muu\sigma+2\kappa\tau+\frac{\muu^2-1}{2\kappa}\tau+\arctan\Big[\muu^{-1}\tanh(\sigma-\frac{\muu}{\kappa}\tau)\Big]\ ,\nonumber\\
2\beta &= \muu\sigma-\frac{\muu^2-1}{2\kappa}\tau+\arctan\Big[\muu^{-1}\tanh(\sigma-\frac{\muu}{\kappa}\tau)\Big]\ ,\nonumber\\
\cos\theta &= \frac{\mathrm{sech}(\sigma-\frac{\muu}{\kappa}\tau)}{\sqrt{1+\muu^2}}\ .\label{422}
\end{align}
These $\beta$ and $\theta$ indeed solve the $q=0$ LL equations: they describe
the known ``pulse'' or ``spin wave'' LL soliton found in
\cite{TjonWright,Takhtajan:1977rv,Fogedby}.\foot{This soliton is
non-topological (i.e. it can be continuously deformed into the vacuum $\theta=
\frac{\pi}{2}$). Upon semiclassical quantization \ci{Jevicki:1978yv,Fogedby}
its $U(1)$ charge $J_2$ is quantized and the quantum soliton $J_2=1$ state may
be identified with the elementary magnon state.} The corresponding conserved
charges are
\begin{align}
E_{_{\rm LL}} & = \frac{\hv}{2}\int\dif x\,\big(\theta'^2+\sin^2 2\theta\, \beta'^2\big)
= \frac{\hv}{\kappa}\ , \la{423}\\
{P}_{_{\rm LL}} & = \hh \int\dif x\,\big(1 + \cos 2\theta\big) \beta' = 2\hh \arctan \muu^{-1}\ ,\la{424} \\
J_\beta & = \hv \int\dif x\, (1 + \cos 2\theta ) = \frac{4\hv\kappa}{1+\muu^2}\ , \la{425}
\end{align}
where we have subtracted the values for the ground state solution
$\theta=\pi/2,\ \phi_1=\kappa\tau,\ \phi_2=0$ to obtain finite expressions.
Here $J_\beta$ is the angular momentum corresponding to translations in
$\beta$.

Comparing with \eqref{gmKappaexpansion2} we see that in the Landau-Lifshitz
limit we have
\begin{align}
{ P}_{_{\rm LL}} = \hh\, \Delta\phi_1
\ , \ \ \ \ \ \ \ \ \ \ \ J_\beta = 2J_2
\ . \label{426}
\end{align}
This then supports the identification of $\Delta \phi_1$ with the spatial
momentum $\pp$ and leads to the following familiar dispersion relation for the
LL soliton
\begin{align}
E_{_{\rm LL}} &= \frac{2\hv^2}{J_2}\sin^2\frac{\pp}{2}
\ , \ \ \ \ \ \ \ \qquad
\pp = \hh^{-1} { P}_{_{\rm LL}} \ , \la{427}
\end{align}
which is also the leading term in the large $J_2$ expansion of the dyonic giant
magnon energy in \rf{dDisp}, $E-J_1\to J_2 + E_{_{\rm LL}}$.\foot{In the $AdS_5
\times S^5$ case the leading term of the expansion is protected and thus it
also agrees with the small $\hh$ expansion of the dyonic giant magnon energy,
matching the expression following from the coherent state expectation value of
the one-loop ferromagnetic spin chain Hamiltonian.}

The generalisation of the relevant large $\k$ expansion of the giant magnon
solution to $q\not=0$ is
\begin{align}
2\alpha &= -\muu(\sigma+q\tau)+2\kappa\tau
+\frac{(1-q^2)(\muu^2-1)}{2\kappa}\tau+\arctan\Big[\muu^{-1}\tanh(\sigma+q\tau-\frac{\muu}{\kappa}(1-q^2)\tau)\Big]\ ,\nonumber\\
2\beta &= \muu(\sigma+q\tau)-\frac{(1-q^2)(\muu^2-1)}{2\kappa}\tau+\arctan\Big[\muu^{-1}\tanh(\sigma+q\tau-\frac{\muu}{\kappa}(1-q^2)\tau)\Big]\ ,\nonumber\\
\cos\theta &= \frac{\mathrm{sech}\big[\sigma+q\tau-\frac{\muu}{\kappa}(1-q^2)\tau\big] }{\sqrt{1+\muu^2}}\ . \la{428}
\end{align}
These $\beta$ and $\theta$ satisfy the $q\ne 0$ LL equations of motion for
\rf{410} while $\alpha$ solves the Virasoro constrains
\begin{align}
\partial_\pm\alpha- C_\pm &= \kappa -\frac{1}{2\kappa} (1- q^2)\,\mathrm{sech}^2\big[\sigma+q\tau-\frac{\muu}{\kappa}(1-q^2)\tau\big]\ .
\end{align}
Note that one can also obtain the $q\ne 0$ solution for $\beta$ and $\theta$ by
applying the worldsheet coordinate transformation $\tilde \tau = \tau$,
$\sigma\rightarrow \tilde \sigma = \sigma - q\tau$, $\partial_\sigma =
\tilde\partial_\sigma$, $\partial_\tau =
\tilde\partial_\tau-q\tilde\partial_\sigma$ after which the LL equations
following from \rf{410} take the standard form $\tilde\partial_\tau n_i =
\frac{1}{2\kappa}(1-q^2)\varepsilon_{ijk}n_j\tilde\partial_\sigma^2n_k.$

The corresponding energy \rf{411} that generalises \rf{423} to the $q\not=0$
case is found to be (taking $\ccc=1$)
\begin{align}
E_{_{\rm LL}} &= \hv\int\dif x\,\Big[-q(1+\cos 2\theta )\beta'+\frac{1}{2}\blambda (\theta'^2+\sin^2 2\theta\, \beta'^2)\Big] \nonumber\\
&= -2\hv q\arctan {\muu}^{-1} + (1-q^2)\frac{\hv}{\kappa}\ , \la{430}
\end{align}
while the expressions for ${P}_{_{\rm LL}}$ and $J_\beta$ remain the same as in
\rf{424} and \rf{425}. As a result, eq. \rf{426} is unchanged while we find
the following generalisation of the LL soliton dispersion relation \rf{427}
\begin{align}
E_{_{\rm LL}} =-q\hv \pp + \frac{2\hv^2(1-q^2)}{J_2}\sin^2\frac{\pp}{2} \ , \ \ \ \ \ \ \ \ \ \ \ \ \ \ \pp= \hh^{-1} P_{_{\rm LL}} \ . \la{431}
\end{align}
This agrees with the large $J_2$ expansion of the giant magnon energy \rf{340}
found in section \ref{sec:giantMagnon} thus supporting the identification of
the magnon momentum \rf{momid} made there.

\section[Symmetry algebra of light-cone gauge S-matrix and exact dispersion relation]{Symmetry algebra of light-cone gauge S-matrix and exact \texorpdfstring{\\}{} dispersion relation}\label{sec:5}

In this section we will go back to the discussion of the world-sheet S-matrix
of the mixed-flux $AdS_3 \times S^3$ theory of
\cite{Hoare:2013pma,Hoare:2013ida}. We shall first review the symmetry algebra
that underlies the light-cone gauge S-matrix. Then we shall suggest a
modification of one of the conjectures in \cite{Hoare:2013ida} to find that we
can recover the semiclassical $q\not=0$ dyonic giant magnon dispersion relation
\rf{340} derived in section \ref{sec:dr} by considering the bound states of the
theory and taking an appropriate strong-coupling limit.

\subsection{Symmetry algebra}

As the type IIB supergravity backgrounds with NS-NS and R-R 3-form fluxes are
related by S-duality, the space-time symmetry of our background should not
depend on $q$. Indeed, the $AdS_3\times S^3$ part of the world-sheet action
can be described by the same supercoset geometry $[PSU(1,1|2) \times
PSU(1,1|2)]/[SU(1,1) \times SU(2)]$ \cite{Pesando:1998wm} with $q$ appearing
only as a parameter in the action \cite{Cagnazzo:2012se}.

For this reason it is not surprising that the symmetry algebra of the
world-sheet S-matrix \cite{Hoare:2013pma,Hoare:2013ida} describing scattering
above the BMN string (which should be a subalgebra of the supercoset symmetry
preserved by the BMN vacuum) should not depend on $q$. The dependence on $q$
then enters through the form of its representation on states
\cite{Hoare:2013ida}.

The relevant symmetry takes the form of a direct sum of two copies of an
algebra with the central extensions identified.\footnote{The symmetry algebra
here is the same as in the case of the S-matrix of the Pohlmeyer-reduced theory
corresponding to the $AdS_3 \times S^3$ superstring
\cite{Hoare:2011fj,Grigoriev:2008jq}.} The generators of a single copy of this
algebra are: $(i)$ two $U(1)$ generators $\mathfrak R$ and $\mathfrak L$;
$(ii)$ four supercharges $\mathfrak Q_{\pm\mp}$ and $\mathfrak S_{\pm\mp}$
($\pm$ denote the charges under the $U(1) \times U(1)$ bosonic subalgebra);
$(iii)$ three central extension generators $\mathfrak C$, $\mathfrak P$ and
$\mathfrak K$. Defining
\begin{equation}
\mathfrak{M} =\frac12 (\mathfrak R + \mathfrak L)\ , \qquad
\mathfrak B =\frac12( \mathfrak R - \mathfrak L)\ ,
\end{equation}
the non-vanishing (anti-)commutation relations are given by
\begin{align}
&[\mathfrak B, \, \mathfrak Q_{\pm\mp}] = \pm i \mathfrak Q_{\pm\mp}\ ,
& \nonumber
&[\mathfrak B, \, \mathfrak S_{\pm\mp}] = \pm i \mathfrak S_{\pm\mp}\ ,
&& \\
&\{\mathfrak Q_{\pm\mp} , \,\mathfrak Q_{\mp\pm}\} = \mathfrak P\ ,
& \label{!bcd}
&\{\mathfrak S_{\pm\mp} , \,\mathfrak S_{\mp\pm}\} = \mathfrak K\ ,
&& \{\mathfrak Q_{\pm\mp} , \,\mathfrak S_{\mp\pm}\} = \pm i\, \mathfrak M + \mathfrak C\ .
\end{align}
These are consistent with the following set of reality conditions
\begin{equation}\label{!bhrc}
\mathfrak B^\dagger = -\mathfrak B\ , \qquad
\mathfrak Q_{\pm\mp}^\dagger = \mathfrak S_{\mp\pm}\ , \qquad
\mathfrak M^\dagger = -\mathfrak M\ ,\qquad
\mathfrak P^\dagger = \mathfrak K\ , \qquad
\mathfrak C^\dagger = \mathfrak C\ .
\end{equation}
This superalgebra is a centrally-extended semi-direct sum of $\mathfrak u(1)$
(generated by $\mathfrak B$) with two copies of the superalgebra
$\mathfrak{psu}(1|1)$, i.e.
\begin{equation}\label{!al}
[\mathfrak u(1)\inplus \mathfrak{psu}(1|1)^2 ] \ltimes \mathfrak u(1) \ltimes \mathbb{R}^3\ .
\end{equation}
The central extensions are represented by the generators $\mathfrak M$
(corresponding to $\mathfrak u(1) $) and $\mathfrak C$, $\mathfrak P$ and
$\mathfrak K$. There is only a single copy of the four central extensions in
the symmetry of the full S-matrix
\begin{equation}\label{!als}
[\mathfrak u(1)\inplus \mathfrak{psu}(1|1)^2]^2\ltimes \mathfrak u(1) \ltimes \mathbb{R}^3\ .
\end{equation}

The particular (reducible) representation of this symmetry algebra of interest
to us here consists of one complex boson $\phi$ and one complex fermion $\psi$.
The action of the $U(1)$ and fermionic generators are discussed in
\cite{Hoare:2013ida}; here we will just focus on the central extensions. These
generators have the following action on the one-particle states
\begin{align}\label{!bhabcd}
\{\mathfrak M, \mathfrak C, \mathfrak P,\mathfrak K\}|\Phi_\pm\rangle
=\{\pm \tfrac i2 M_{_\pm},C_{_\pm},P_{_\pm},K_{_\pm}\}|\Phi_\pm\rangle\ ,
\end{align}
where $\Phi_\pm \in \{\phi_\pm,\psi_\pm\}$. These representation parameters
should be real functions of the energy and momentum of the state. Furthermore,
the closure of the algebra requires that these four parameters satisfy a
constraint that is interpreted as the dispersion relation
\begin{equation}\label{!dr}
C_{_\pm}^2 = \frac{M_{_\pm}^2}{4} + P_{_\pm} K_{_\pm}\ .
\end{equation}

The tree-level S-matrix was computed in \cite{Hoare:2013pma} and from this
result the leading-order expressions for the representation parameters in the
near-BMN expansion ($\textrm{h} \to \infty$) were written down in
\cite{Hoare:2013ida}:
\begin{equation}
M_{_\pm} = 1\pm q \, \py\ , \qquad
C_{_\pm} =\, \frac{\energy_{_\pm}}{2}\ , \qquad
P_{_\pm} = - \frac{i}{2} \, \sqrt{1-q^2} \, \py\ , \qquad
K_{_\pm} = \frac{i}{2}\, \sqrt{1-q^2} \, \py\ .\la{58}
\end{equation}
Here, the momentum $p$ of a near-BMN excitation is related to the magnon
momentum $\pp$ of sections \ref{sec:giantMagnon} and \ref{sec:llLimit} in the
usual way, i.e. through a rescaling by the string tension $\textrm{h}$
\begin{equation}
\py = \frac{\pp}{\textrm{h}}\ .
\end{equation}
Substituting these near-BMN expressions \rf{58} into \eqref{!dr} we reproduce
the expected near-BMN dispersion relation \cite{Hoare:2013pma}
\begin{equation}\label{!nbmndr}
\energy_{_\pm} = \sqrt{(1 \pm q \, \py)^2 + (1-q^2)\py^2} = \sqrt{1-q^2 + (p \pm q)^2}\ .
\end{equation}

Exact completions of $C_\pm,P_\pm$ and $K_\pm$ were then proposed in
\cite{Hoare:2013ida} based on various algebraic requirements and analogy with
the pure R-R flux case:
\begin{equation}\label{!exact}
C_{_\pm} =\, \frac{\energy_{_\pm}}{2}\ , \qquad
P_{_\pm} = \tfrac{\textrm{h}}2 \sqrt{1-q^2}(1-e^{i\pp})\ , \qquad
K_{_\pm} = \tfrac{\textrm{h}}2 \sqrt{1-q^2}(1-e^{-i\pp})\ .
\end{equation}
Substituting these exact expression into the dispersion relation \eqref{!dr} we
find
\begin{equation}
\energy_{_\pm} = \sqrt{M_{_\pm}^2 + 4 \textrm{h}^2 (1-q^2) \sin^2 \frac{\pp}{2}} \ .\la{512}
\end{equation}
An exact completion for $M_{_\pm}$ conjectured in \cite{Hoare:2013ida} was
\be\la{512a} M_{_\pm} = 1
\pm 2 q \textrm{h} \sin \frac{\pp}{2} \ . \ee
However, it is now clear that this is not consistent with the semiclassical
result \eqref{340}. Instead, the expression that is consistent with both the
near-BMN limit \eqref{!nbmndr} and the semiclassical result \eqref{340} is
simply
\begin{equation}
M_{_\pm} = 1 \pm q \textrm{h} \pp \ . \la{513}
\end{equation}
This alternative completion is not only compatible with the semiclassical
result, but also has another advantage in that the construction of the
bound-state dispersion relation is far more natural than in the case of
\rf{512a}, as we will explain below.

\subsection{Zhukovsky variables}

To discuss the bound-state dispersion relation we need to briefly describe the
effect of the choice of \rf{513} as oppose to \rf{512a} on the exact S-matrix.
In \cite{Hoare:2013ida} the S-matrix was written down in terms of the Zhukovsky
variables $x^\pm_{_\pm}$ (to be defined below), up to four overall phase
factors. Various equations for these overall factors that follow from
unitarity, braiding unitarity and crossing symmetry were listed there. Here we
will not address the issue of these factors, which is still an open question
for the mixed-flux case, i.e. will leave them unfixed.\footnote{For the pure
R-R case there have been a number of works studying these overall factors
\cite{Abbott:2012dd,Beccaria:2012kb,Sundin:2013ypa,Abbott:2013mpa}, leading to
a conjecture in \cite{Borsato:2013hoa}. In \cite{Hoare:2013ida} there was some
discussion of the strong-coupling limit of these factors in the mixed-flux
case. However, this discussion is likely to need modification in light of the
results of this paper, and furthermore, the recipe presented there cannot be
extend beyond this strong-coupling limit, due to inconsistencies with unitarity
\cite{Engelund:2013fja}. We thank R.~Roiban for drawing this last point to our
attention.}

The key point is that the S-matrix expressed as a function of the Zhukovsky
variables $x_{_\pm}^\pm$ maintains exactly the same form as in
\cite{Hoare:2013ida}. It is only the map from $x_{_\pm}^\pm$ to the
energy/momentum of the scattering states and their dispersion relation that are
modified due to the change from \rf{512a} to \rf{513}. It is important to note
that many of the key properties of the S-matrix, including the Yang-Baxter
equation, unitarity, braiding unitarity and crossing symmetry are satisfied (so
long as the overall factors satisfy the equations as written in
\cite{Hoare:2013ida}) without the use of either this map or the dispersion
relation.

The map between the Zhukovsky variables and the energy/momentum corresponding
to \rf{512} with arbitrary $M_\pm$ is a straight-forward generalisation of the
familiar one:\footnote{Here the definitions of $x_{_\pm}^\pm$ have a natural
periodic extension of the region $\pp\in (0, \pi)$ to the whole line, which is
consistent with the semiclassical identification of $\pp$ with the angular
separation of the dyonic giant magnon string end-points.}
\begin{equation}\begin{split}\label{!map}
& e^{i\pp} = \frac{x_{_\pm}^+}{x_{_\pm}^-}\ , \qquad
\energy_{_\pm} = \frac{\textrm{h}\sqrt{1-q^2}}{2i} \big(x^+_{_\pm} - \frac{1}{x^+_{_\pm}} - x^-_{_\pm} + \frac{1}{x^-_{_\pm}}\big)\ ,
\\
& x_{_\pm}^\pm = r_{_\pm} e^{\pm i\frac{\pp}{2}}\ , \qquad
r_{_\pm} = \frac{\energy_{_\pm} + M_{_\pm}}{2\textrm{h}\sqrt{1-q^2} \,\sin\frac{\pp}{2}} = \frac{2\textrm{h}\sqrt{1-q^2}\,\sin\frac{\pp}{2}}{\energy_{_\pm} - M_{_\pm}}\ ,
\end{split}\end{equation}
with the dispersion relation \rf{512} expressed as
\begin{equation}\label{!drxpm}
x^+_{_\pm} + \frac{1}{x^+_{_\pm}} - x^-_{_\pm} - \frac{1}{x^-_{_\pm}} = \frac{2 i M_{_\pm}}{\textrm{h}\sqrt{1-q^2}}\ .
\end{equation}
As discussed above, the $M_{_\pm}$ is unconstrained by the algebra, and hence
could be an arbitrary function of the momentum provided it has the required
near-BMN and strong-coupling limits. The choice of \rf{513} corresponds to
\begin{equation}\label{!defm}
M_{_\pm} = 1 \pm q \textrm{h} \pp = 1 \mp i q \textrm{h} \log \frac{x_{_\pm}^+}{x_{_\pm}^-}\ .
\end{equation}
Substituting \rf{!defm} for $M_{_\pm}$ in \eqref{!drxpm} we get
\begin{equation}\label{!curve}
\Big[\sqrt{1-q^2}\big(x^+_{_\pm} + \frac{1}{x^+_{_\pm}}\big) \mp 2 q \log x_{_\pm}^+\Big] - \Big[\sqrt{1-q^2}\big(x^-_{_\pm} + \frac{1}{x^-_{_\pm}}\big) \mp 2 q \log x_{_\pm}^-\Big]
= \frac{2 i}{\textrm{h}}\ .
\end{equation}
It follows from this representation that we can define a ``generalised''
Zhukovsky map
\begin{equation}\label{!zmap}
\sqrt{1-q^2}\big(x_{_\pm} + \frac{1}{x_{_\pm}}\big) \mp 2 q \log x_{_\pm} = u \ ,\qquad \qquad x^\pm_{_\pm} = x_{_\pm}(u \pm \frac i{\textrm{h}})\ ,
\end{equation}
that ``solves'' the dispersion relation. However, it is apparent that the
analytic structure of the inverse, $x_{_\pm}(u)$, of \eqref{!zmap} will be
considerably more complicated than in the case of the pure R-R flux ($q=0$),
and indeed it is not clear whether this set of variables is the most
illuminating for studying the complex structure of the spectral curve
\eqref{!curve}.

\subsection{Bound-state dispersion relation}

To construct the bound-state dispersion relation we first need to know the
position of the poles of the S-matrix. Technically, for this we should also
have the exact form of the four phase factors mentioned above. However, at the
position of a pole corresponding to a bound state we expect that the residue of
the S-matrix given in \cite{Hoare:2013ida} should project onto a short
representation of the symmetry algebra (i.e. a boson and a fermion). From the
form of the S-matrix it is then clear that candidate positions for poles (and
the corresponding short bound-state representations) include\footnote{Here the
unprimed and primed variables correspond to the two incoming particles in the
scattering process.}
\begin{equation}\begin{split}\label{!bspoles}
(i): & \quad x^+_{_\pm} = x^{\prime-}_{_\pm}\ , \qquad
\big\{|\phi_\pm\phi^\prime_\pm\rangle ,
\ |\phi_\pm\psi^\prime_\pm\rangle + \varphi_{_\pm}\varphi^\prime_{_\pm}\frac{\eta_{_\pm}}{\eta^\prime_{_\pm}} |\psi_\pm \phi^\prime_\pm\rangle \big\}\ ,
\\
(ii): & \quad x^-_{_\pm} = x^{\prime+}_{_\pm}\ , \qquad
\big\{|\psi_\pm\psi^\prime_\pm\rangle ,
\ |\phi_\pm\psi^\prime_\pm\rangle - \varphi_{_\pm}\varphi^\prime_{_\pm}\frac{\eta_{_\pm}}{\eta^\prime_{_\pm}} |\psi_\pm \phi^\prime_\pm\rangle \big\}\ ,
\end{split}\end{equation}
where
\begin{equation}
\varphi_{_\pm} = \sqrt[4]{\frac{x^+_{_\pm}}{x^-_{_\pm}}}\ ,\qquad\qquad
\eta_{_\pm} = \sqrt{i(x^-_{_\pm} - x^+_{_\pm})}\ . \qquad
\end{equation}

On physical grounds we would expect that the bound states should form in the
sector associated to the 3-sphere and indeed this is what happens in the $q=0$
case of pure R-R flux \cite{Borsato:2013qpa}. As the field $\phi_\pm$ is
associated to the 3-sphere we expect there to be a pole corresponding to case
$(i)$ in \eqref{!bspoles}, and not to case $(ii)$.

The bound-state energy and momentum should be given by the sum of those of the
two constituent states
\begin{equation}
E^{(2)}_{_\pm} = \energy_{_\pm} + \energy^\prime_{_\pm}\ , \qquad \qquad \pp^{(2)} = \pp + \pp^\prime\ .
\end{equation}
From \eqref{!map} it immediately follows that
\begin{equation}
e^{i\pp^{(2)}} = \frac{x^{\prime+}_{_\pm}}{x^-_{_\pm}}\ , \qquad
E^{(2)}_{_\pm} = \frac{\textrm{h}\sqrt{1-q^2}}{2i} \big(x^{\prime+}_{_\pm} - \frac{1}{x^{\prime+}_{_\pm}} - x^-_{_\pm} + \frac{1}{x^-_{_\pm}}\big)\ .
\end{equation}
Therefore, we can interpret the Zhukovsky variables for the bound state as
\begin{equation}
x^{(2)+}_{_\pm} = x_{_\pm}^{\prime+}\ , \qquad
x^{(2)-}_{_\pm} = x_{_\pm}^{-}\ .
\end{equation}
Denoting $x^+_{_\pm} = x'^-_{_\pm} = \textrm{x}_{_\pm}$ at the position of the
pole, the dispersion relations for the two constituent particles of the bound
state are then
\begin{equation}
\textrm{x}_{_\pm} + \frac{1}{\textrm{x}_{_\pm}} - x^{(2)-}_{_\pm} - \frac{1}{x^{(2)-}_{_\pm}} = \frac{2 i M_{_\pm}}{\textrm{h}\sqrt{1-q^2}}\ ,\qquad
x^{(2)+}_{_\pm} + \frac{1}{x^{(2)+}_{_\pm}} - \textrm{x}_{_\pm} - \frac{1}{\textrm{x}_{_\pm}} = \frac{2 i M^\prime_{_\pm}}{\textrm{h}\sqrt{1-q^2}}\ .
\end{equation}
Summing these up we find the dispersion relation for the bound state
\begin{equation}
x^{(2)+}_{_\pm} + \frac{1}{x^{(2)+}_{_\pm}} - x^{(2)-}_{_\pm} - \frac{1}{x^{(2)-}_{_\pm}} = \frac{2 i M^{(2)}_{_\pm}}{\textrm{h}\sqrt{1-q^2}}\ ,
\end{equation}
where
\begin{equation}
M^{(2)}_{_\pm} = M_{_\pm} + M^\prime_{_\pm}\ .
\end{equation}
We thus see that the eigenvalue $M_{_\pm}$ of the central generator
$\mathfrak{M}$ is additive when acting on the two (and higher) particle states.
Indeed, this follows immediately from the fact that its coproduct is the
standard one \cite{Hoare:2013ida,Borsato:2013qpa,Hoare:2013pma,Borsato:2012ud}
(i.e. it is just given by the usual Leibniz action).

It follows from the definition of $M_{_\pm}$ in \eqref{!defm} that the value of
$\mathfrak{M}$ acting on the bound state is
\begin{equation}
M^{(2)}_{_\pm} = 2 \pm q\textrm{h}\pp^{(2)} = 2 \mp i q \textrm{h} \log\frac{x^{(2)+}_{_\pm}}{x^{(2)-}_{_\pm}}\ .
\end{equation}
It is crucial to note that the simplicity found here, in particular, the fact
that $\textrm{x}^\pm$ drop out without any additional work, is a direct
consequence of the fact that $M_{_\pm}$ in \rf{513},\rf{!defm} are {\it linear}
functions of $\pp$. Furthermore, it is clear that this procedure will iterate,
giving a tower of bound states with
\begin{equation}
M^{(\Nb)}_{_\pm} = N \pm q\textrm{h}\pp^{(\Nb)} = N \mp i q \textrm{h} \log\frac{x^{(\Nb)+}_{_\pm}}{x^{(\Nb)-}_{_\pm}}\ .
\end{equation}
The resulting dispersion relation for the $N$-particle bound state is then
\begin{equation}\la{530}
E^{(\Nb)}_{_\pm} = \sqrt{\big(N \pm q\textrm{h}\pp^{(\Nb)}\big)^2 + 4\textrm{h}^2(1-q^2)\sin^2 \frac{\pp^{(\Nb)}}{2}} \ .
\end{equation}
This agrees with the semiclassical result \eqref{340} after quantizing the
angular momentum: $J_2 = N$.

\section{Concluding remarks}\label{sec:6}

In this paper we have supplemented the information provided by the
perturbative near-BMN expansion \cite{Hoare:2013pma} and the light-cone
symmetry algebra \ci{Hoare:2013ida} with the construction of the
semiclassical dyonic giant magnon solution in $AdS_3 \times S^3\times T^4$
string theory with mixed flux to propose the exact form of the corresponding
dispersion relation. We have seen that the presence of the WZ term
representing the NS-NS flux in the bosonic string action leads to subtleties
associated to the proper choice of boundary terms and the definition of angular
momenta, which become important for non-trivial open-string solutions like the
giant magnon.

We reviewed the symmetry algebra for the string light-cone gauge S-matrix
and introduced a new set of Zhukovsky variables corresponding to the proposed
dispersion relation. Analyzing the resulting bound-state dispersion
relation, we found that it has a simple structure \rf{530} and agrees with
the giant magnon dispersion relation \rf{340}. The implications of this new
dispersion relation for the structure of the yet undetermined ``phase factors''
in the exact S-matrix \ci{Hoare:2013ida} remain to be studied.

It would be interesting to provide further checks of the dispersion relation
\rf{5},\rf{7}. One possibility would be through a two-loop perturbative string
computation like that done in the $AdS_5 \times S^5$ \ci{Klose:2007rz} and
$AdS_2 \times S^2\times T^6$ cases \ci{Murugan:2012mf}. It appears, however,
that the near-flat space limit \ci{Maldacena:2006rv} used in these papers is
not sufficient to determine the $q$-dependence of the two-loop correction to
the dispersion relation (e.g., a potential $qp^3$ term will not be seen in this
limit). Therefore, to check \rf{7} one would need to do the full near-BMN
two-loop computation of the two-point function, which is yet to be performed in
the $AdS_5 \times S^5$ case. One may also get additional information about the
perturbative expansion of the dispersion relation and S-matrix using
unitarity-based methods \ci{Bianchi:2013nra,Engelund:2013fja}.

Another check of \rf{5},\rf{7} would be to confirm that the first semiclassical
(one-loop) correction to the giant magnon energy \rf{340} vanishes\foot{In
semiclassical limit $J_2 \sim h {\cal J}_2 $ and ${\cal J}_2 $ and $\pp$ are
fixed while one expands in large $\hh$.} as was shown in the case of the $AdS_5
\times S^5$ giant magnon in
\cite{Minahan:2006bd,Papathanasiou:2007gd,Chen:2007vs}. This should indeed be
the case since (i) the one-loop corrections in the string and the corresponding
Pohlmeyer reduced theory should match \ci{Hoare:2009rq} (since the classical
equations and thus the leading fluctuations near a classical solution are
directly related) and (ii) the solution of the reduced theory corresponding to
the giant magnon is essentially the same as in the $q=0$ case up to a simple
rescaling of the mass scale by $\sqrt{1-q^2}$ (see \ci{Hoare:2013ida} and the
appendix below).

Finally, to use the dispersion relation \rf{5},\rf{7} and the corresponding
S-matrix as a starting point for computing the string spectrum, it would be
important to have a better understanding of the analytic structure of the
complex spectral curve \eqref{!curve}, in particular, identifying the
corresponding uniformizing variables (the analogs of those introduced in the
$AdS_5 \times S^5$ case in \cite{Janik:2006dc,Arutyunov:2007tc}). 

\section*{Acknowledgments}

We would like to thank S. Frolov, O. Ohlsson-Sax, R. Roiban and L. Wulff for 
useful discussions. BH is supported by the Emmy Noether Programme ``Gauge
fields from Strings'' funded by the German Research Foundation (DFG). AS and
AAT acknowledge the support of the STFC grant ST/J000353/1. The work of AAT is
also supported by the ERC Advanced grant No.290456.

\appendix

\section{Relation to soliton of the Pohlmeyer reduced theory}\label{app:a}

Let us briefly describe the relation between the $q\not=0$ generalisation of
the giant magnon solution, found in section \ref{sec:32}, and the corresponding
soliton of the Pohlmeyer reduction of $\mathbb{R} \times S^3$ string theory
with $q\not=0$, which is the complex sine-Gordon model with the mass parameter
rescaled by $\sqrt{1-q^2}$ \cite{Hoare:2013pma}. This generalises the relation
between the soliton of the complex sine-Gordon model and the $q=0$ dyonic giant
magnon used in \cite{Chen:2006gea}.

The Lagrangian density of the Pohlmeyer reduced theory and the relation of the string
embedding coordinates $X_m$ to the reduced variables are given by \cite{Hoare:2013pma}
\begin{align}
\lagr & = \partial_+\varphi\partial_-\varphi+\tan^2\varphi\, \partial_+\chi\partial_-\chi+\frac12\kappa^2(1-q^2)\cos2\varphi\ ,\nonumber\\
\kappa^2\cos2\varphi & = \partial_+X\cdot\partial_-X\ ,
\qquad \kappa^3\sin^2\varphi\, \partial_\pm\chi = \mp\frac{1}{2}\varepsilon_{mnkl}X^m\partial_+X^n\partial_-X^k\partial_\pm^2X^l\ .\label{PRvars}
\end{align}
These can be written in terms of the $SU(2)$ current $\alg{J}$ in \eqref{jparam}
as follows:
\begin{align}
\kappa^2\cos 2\varphi = -\frac{1}{2}\tr(\alg{J}_+\alg{J}_-)\ ,
\qquad \kappa^3\sin^2\varphi\, \partial_\pm\chi = \pm\frac{1}{8}\tr\big([\alg{J}_+,\alg{J}_-]\partial_\pm \alg{J}_\pm\big)\ .\label{PRvarsFR}
\end{align}
Substituting the expressions \rf{323}--\rf{327} for the $q\not=0$ giant magnon solution
into $\alg{J}_\pm$ the corresponding reduced theory solution is found to be
\begin{align}
\sin\varphi = \frac{\cos\al}{\cosh\big[\cos\al({\cal X}+q{\cal T})\big]}\ ,\qquad\qquad
\chi = 2\sin\al ({\cal T}+ q{\cal X})\ ,
\end{align}
where ${\cal T}$ and ${\cal X} $ were defined in \rf{32}.
Then
\begin{align}
\psi \equiv \sin\varphi \, e^{ \frac i2 \chi}
= \frac{\cos\al\, \exp\big[i\sin\al \, ({\cal T}+ q{\cal X})\big]}{\cosh\big[\cos\al\, ({\cal X}+q{\cal T})\big]} 
\end{align}
is recognised as the familiar complex sine-Gordon soliton solution.

\parskip=0.pt
\baselineskip 11pt

\bibliographystyle{utphysM}
\bibliography{HST}

\end{document}